\documentclass[showpacs,preprintnumbers,amsmath,amssymb]{revtex4}

\usepackage{graphicx}
\usepackage{dcolumn}
\usepackage{bm}

\newcommand{\beq}{\begin{equation}}
\newcommand{\beqa}{\begin{eqnarray}}
		  \newcommand{\eeq}{\end{equation}}
\newcommand{\eeqa}{\end{eqnarray}}

\newcommand{\lsim}{\lesssim}

\newcommand{\lmk}{\left(}
\newcommand{\rmk}{\right)}

\newcommand{\lla}{\left\langle}

\newcommand{\rra}{\right\rangle}
\newcommand{\so}{M_\odot}

\begin{document}

\preprint{KUNS-2318}

\title{ Detector configuration of  DECIGO/BBO and  identification of cosmological neutron-star binaries}
\author{Kent Yagi and Naoki Seto}
\affiliation{
Department of Physics, Kyoto University,
   Kyoto, 606--8502, Japan
}

\date{\today}

\begin{abstract}
The primary target for the  planned space-borne gravitational wave interferometers DECIGO/BBO is a primordial gravitational wave background (PGWB).
However there exist astrophysical foregrounds and among them, gravitational waves from neutron star (NS) binaries are the solid and strong component that must be identified and subtracted.
In this paper, we discuss the geometry of detector configurations preferable for identifying  the NS/NS binary signals.
As a first step, we analytically estimate the minimum signal-to-noise ratios (SNRs) of the binaries for several  static detector configurations that are characterized by adjustable geometrical parameters, and determine the optimal values for these parameters.
Next we perform numerical simulations to take into account the effect of detector motions, and find  reasonable agreements with the analytical results.
We show that, with the standard network formed by 4 units of triangle detectors,  the proposed BBO sensitivity would be sufficient in receiving gravitational waves from all the NS/NS binaries at $z\le 5$ with SNRs higher than 25.  
We also discuss the minimum sensitivity of DECIGO required for the foreground identification.

\end{abstract}

\pacs{Valid PACS appear here}
\maketitle

\section{INTRODUCTION}

Currently, several ground-based gravitational wave interferometers ({\it e.g.} LIGO, VIRGO, GEO and TAMA) are operating or in installation/commissioning phases for upgrades. In the next decade, second generation detectors ({\it e.g.} advanced-LIGO, advanced-Virgo and LCGT) would be available.
These have optimal sensitivities around 100-1000 Hz  aiming to detect gravitational waves (GWs) from NS/NS inspirals, supernovae, etc. Furthermore, the possibilities of more powerful third generation detectors ({\it e.g.} ET) have been actively discussed with primordial gravitational wave background (PGWB) as one of their principle targets, and they would newly explore the lower frequency regime down to $\sim1$ Hz.
Meanwhile, NASA and ESA are planning to launch a space-borne gravitational wave interferometer called the Laser Interferometer Space Antenna (LISA)~\cite{danzmann} whose optimal band is  $0.1-100$ mHz.
GWs from white dwarf (WD) binaries are guaranteed targets for LISA, but they will also form  confusion noises that might mask other signals including PGWB (see Refs.~\cite{allen,maggiore} for reviews on this source).

The Deci-Hertz Interferometer Gravitational Wave Observatory (DECIGO) is a future plan of Japanese space mission for observing GWs around $f\sim 0.1-10$ Hz~\cite{decigo,kawamura2006,kawamura2008}.  In US, the Big Bang Observer (BBO) {has been} proposed as a follow-on mission to LISA, and its optimal band is similar to that of DECIGO~\cite{phinneybbo} (see also Ref.~\cite{atomic} for a proposal of an atomic interferometer).
Since GWs from WD/WD binaries have cut off frequencies at $f\lsim 0.2$Hz~\cite{farmer}, the higher frequency part of DECIGO/BBO is free from their confusion noises.  
Therefore,  the direct detection of the PGWB produced during inflation has been set as the primary goal of DECIGO and BBO.

 For  the amplitude of PGWB generated by standard slow-roll inflation, the WMAP team has placed a conservative bound  corresponding to  $\Omega_{\mathrm{GW}}\lesssim 10^{-14}$ around 1 Hz~\cite{smith}, where $\Omega_{\mathrm{GW}}(f)$ represents the normalized energy density of GW background at a frequency $f$. However, with theoretical analysis based on the temperature anisotropies reported by WMAP, we can reasonably expect a level $\Omega_{\mathrm{GW}}\lsim 10^{-15}$ for the inflation background at the DECIGO/BBO band. 
Since the amplitude of the target PGWB is expected to be  small, the correlation analysis  would be a powerful technique for its detection~\cite{flanagan-corr,allenromano}. The sensitivity of this method is characterized by the so-called overlap reduction function that depends strongly on the geometry of the detector network. It becomes maximum for a pair of co-aligned detectors. With DECIGO and BBO, two units of aligned detectors in the star-of-David form are planned to be used for correlation analysis with their detection limit $\Omega_{\mathrm{GW}}\sim10^{-15}$-$10^{-16}$. In addition to these co-aligned ones, two other units might be launched, mainly to improve the angular resolution of astrophysical sources, by forming long-baseline ($ \sim$1 AU) interferometers (see Fig.~\ref{default}).

While the frequency regime of DECIGO and BBO would not be fundamentally limited by WD/WD binaries, NS/NS,
NS/BH and BH/BH binaries (BH: black hole)  are  considered as their important astrophysical sources.
Among them, the foreground GWs by  NS/NS binaries are understood relatively well with their cosmological merger rate ($10^5$/yr) estimated observationally~\cite{kalogera,cutlerharms}.  Its total amplitude is expected to be $\Omega_{\mathrm{GW}}\sim 10^{-12}$~\cite{cutlerharms}.  Therefore, for {successful operations} of DECIGO and BBO {in} opening deep windows of GWs around 1 Hz, it is essential to identify NS/NS (also NS/BH and  BH/BH) binaries and remove their contributions from data streams of detectors by $\sim$4-5 orders of magnitude in terms of $\Omega_{\mathrm{GW}}$.

Cutler and Harms~\cite{cutlerharms} have done a pioneering work on this topic.
They set the threshold signal-to-noise ratios (SNRs) $\rho_{\mathrm{thr}}$ for binary detection and self-consistently estimated the fraction of binaries whose SNRs are below $\rho_{\mathrm{thr}}$.
This corresponds to the fraction of residual NS/NS binaries that cannot be subtracted.
According to their calculation, with the base design sensitivity of BBO, it is possible to subtract out NS/NS binary signals more than 5 orders of magnitude.
However, if it is worse by a factor of 2, the prospect crucially depends on the value of $\rho_{\mathrm{thr}}$.
If the sensitivity is worse by a factor of 4, then it seems difficult to clean  the foreground noises down to the level needed for the detection of the inflation background.
Harms \textit{et al.}~\cite{harms} have performed numerical simulations of a  projection method for reducing the subtraction noises and confirmed the results obtained by Cutler and Harms~\cite{cutlerharms}.

Our basic aim in this paper is to specifically study how the performance of the binary identification depends on the geometry of the detector network of DECIGO or BBO. 
Here we should comment on some of the interesting aspects on the detector configuration. 
If there is no astrophysical foreground, in order to detect PGWB, it is most favored to put all the detectors co-aligned at the same location so that the  correlation analysis would work for  all the pairs of detectors.
However, in order to identify NS/NS binaries and subtract them,  detectors should have different orientations to cover up the whole sky.
This is particularly relevant for a binary at a high redshift. Such a binary goes through the DECIGO/BBO band in a short amount of time ({\it e.g.} $\sim$1 month for a NS/NS binary at $z\sim5$), and we need multiple detectors with different orientations not to miss its detection.  Thus, determining each configuration of DECIGO and BBO is  an important topic closely related to their primary  scientific goal.
Since the design parameters of DECIGO are currently under discussion \cite{kawa}, another goal of this paper is to clarify its  sensitivity required to sufficiently identify the  NS/NS binaries. 

We consider several examples of possible detector networks by introducing  geometrical parameters that should  be adjusted to maximize the minimum value of the signal-to-noise ratio (SNR) of the binaries. The minimum value would be a helpful reference to discuss the prospect for the identification of all the foreground binaries, and we heavily rely on this reference.  In addition to simple but workable analytical evaluations for the minimum SNRs {with static detector networks},  we include the effect of detector motions by performing  Monte Carlo simulations with the code developed in Refs.~\cite{kent,kent2}.
Then we discuss the  detector sensitivity required for identifying all the NS/NS binaries.
Our analysis might also provide useful insights on the geometrical properties of ground-based detector network.

In this paper, we only discuss the GW foreground made by  NS/NS binaries, as a concrete example. We can expect that BH/NS or BH/BH binaries would {cause} less severe problems,  because of their larger chirp masses ({\it i.e.} larger signals) and their merger rates presumed to be smaller than NS/NS binaries.

This paper is organized as follows.
In Sec.~\ref{sec-decigo}, we briefly describe GW observation with DECIGO and BBO.
In Sec.~\ref{sec-GWB}, we discuss the basic properties of GW foreground of NS/NS binaries.
In Sec.~\ref{sec-det}, we study the responses of  detectors to incoming GW from  a binary.
In Sec.~\ref{sec-dead}, we explain the dead angles for GW interferometers and analytically calculate  the minimum SNRs for various static detector configurations.
In Sec.~\ref{sec-num}, we first explain the set ups for our Monte Carlo simulations.
Then, we provide the numerical  results and their interpretations.
In Sec.~\ref{sec-conclusion}, we summarize our work, suggesting the preferable configurations for DECIGO and BBO, and comment on possible future works.
In Appendix~\ref{app}, we provide formulae that would be useful when relating angular parameters in a spatially fixed frame with the ones in a rotating frame attached to a detector. 

We take the unit $G=c=1$ throughout this paper.

\section{\label{sec-decigo}DECIGO/BBO}

\subsection{Two effective interferometers}

\begin{figure}[tbp]
  \includegraphics[scale=.7,clip]{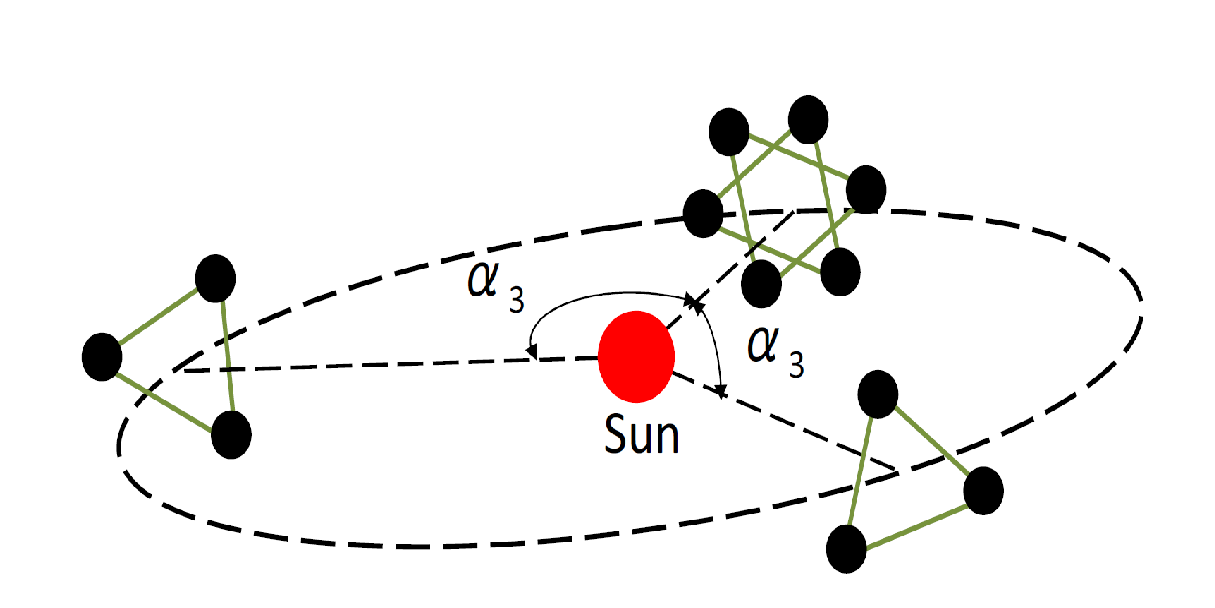} 
 \caption{\label{default} The default configuration of BBO or DECIGO with $\alpha_3=120^\circ$. Totally four units of triangle-like detectors will be operated. Two of them are nearly co-aligned to form a star-of-David constellation for the correlation analysis. Two outrigger ones are used to improve  localization  of individual astrophysical sources. }
\end{figure}

In its original proposal, BBO consists of four units of detectors.  Each unit is composed of three drag-free spacecrafts to form  a nearly  regular  triangle.
Its default configuration is shown in Fig.~\ref{default} with $\alpha_3=120^{\circ}$~\cite{phinneybbo}. These four units  are tilted  60$^{\circ}$ inwards relative to the ecliptic plane to keep its arm lengths nearly constant (as for LISA), and  move  around the sun with the orbital period of 1 yr.  A similar configuration is supposed to be adopted for DECIGO.  The main difference between these two  is the basic design of detectors.  While DECIGO is planned to be Fabry-Perot type interferometers,  BBO will be transponder-type interferometers like LISA.

\begin{figure}[tbp]
  \includegraphics[scale=.6,clip]{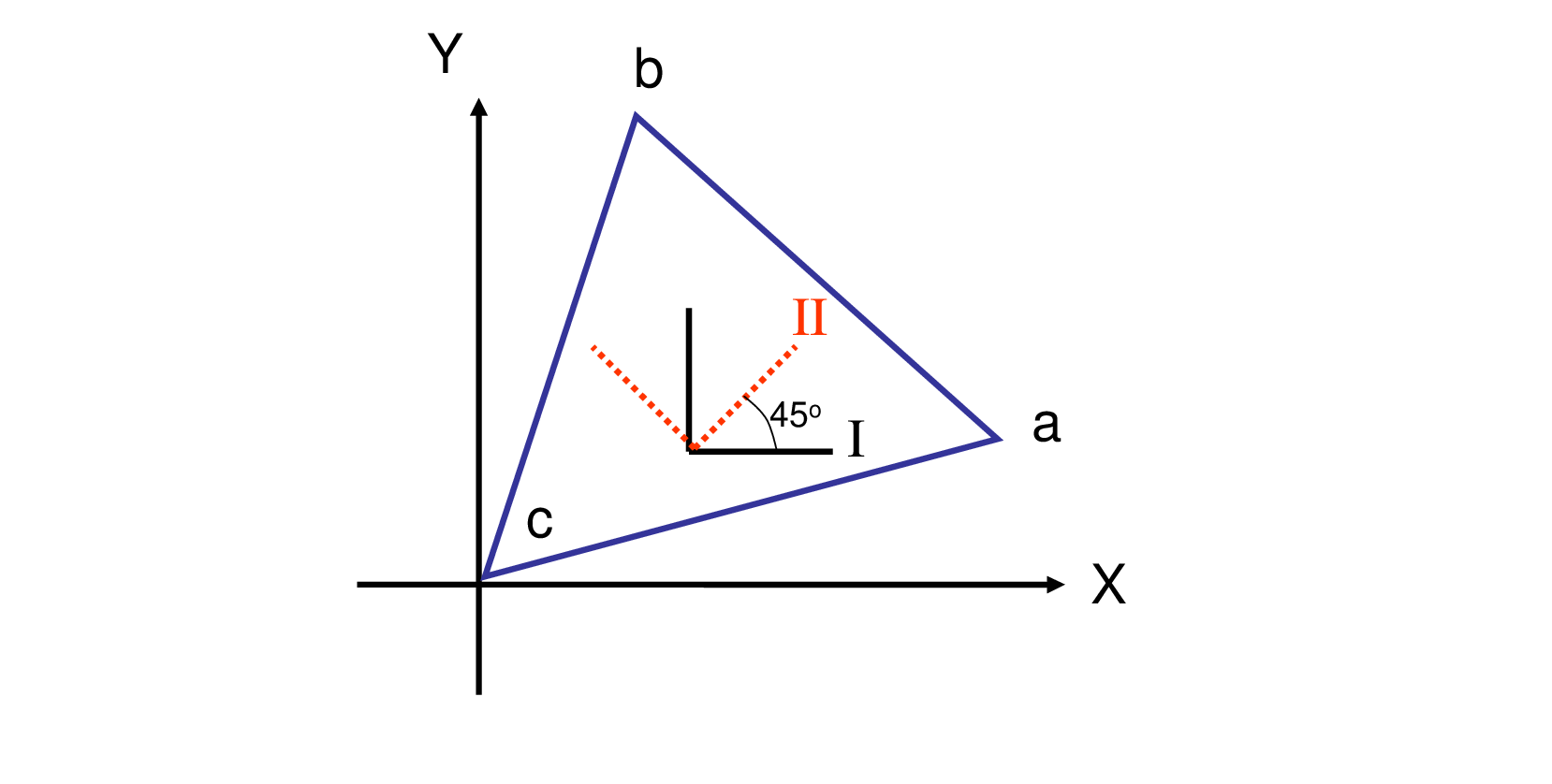} 
 \caption{\label{ae} One unit of BBO (DECIGO) detector and  the orientations of two effectively L-shaped interferometers I and II defined in Eq.~(\ref{tdi}).}
\end{figure}

Here, we briefly sketch a rough idea to generate independent data streams  from each triangle-like unit of BBO  (see {\it e.g.} Ref.~\cite{prince} for a detailed analysis with LISA). 
First we make three interferometers  $a$, $b$ and $c$ defined at the three vertices by using their adjacent two arms (with opening angles $60^\circ$, see Fig.~\ref{ae}) symmetrically. 
Since they share same arms of interferometers, noises in the data from these three interferometers have correlations.
Here, we assume that the covariance matrix of these noises has a simple symmetric structure. Namely, its diagonal elements are the same, and the off-diagonal elements have an identical value. 
Then, in order to obtain the orthogonal data streams,  we take the linear combinations I and II as follows;
\begin{equation}
\mathrm{I}=\frac{a+b-2c}{\sqrt{6}},~~\mathrm{II}=\frac{a-b}{\sqrt{2}}.\label{tdi}
\end{equation}
These essentially correspond to the $A$ and $E$ modes of the TDI data streams in Ref.~\cite{prince}. The $T$ mode is not important for our study.
 Due to the symmetry of the noise matrix of the original data $a$, $b$ and $c$,
the noises of the combined  data I and II have an identical spectrum with no covariance between them \cite{prince,corbin,cutler1998,Seto:2005qy}\footnote{Unless the noises are purely Gaussian distributed, this does not mean that they are statistically  independent.}.
In the long-wave limit, responses of I and II to GWs can be effectively regarded as those of two L-shaped interferometers whose orientations are shown in Fig.~\ref{ae}~\cite{cutler1998,seto-annual}.  
The interferometer II is obtained by $45^\circ$ rotation of the interferometer I.  
Note that the I-II pair forms an orthogonal basis for L-shaped interferometers on the detector plane.
In this paper, we apply the above arguments also for each (triangle-like) unit of DECIGO, assuming that the noise spectra of its two effective interferometers are identical and uncorrelated.  With the four units of triangles  as shown in Fig.~\ref{default}, the total number of the effective (L-shaped) interferometers becomes eight.

\subsection{\label{sec-noise}DECIGO/BBO noise spectrum}

 In this subsection, we provide the noise spectrum of DECIGO and BBO for each effective interferometer mentioned in the previous subsection. 
The data stream $s_\alpha(t)$ of an interferometer $\alpha$ ({\it e.g.} $\alpha=$I or II) can be decomposed into GW signal $h_\alpha(t)$ and the detector noise $n_\alpha(t)$ as
\beq
s_\alpha(t)=h_\alpha(t)+n_\alpha(t).
\eeq
 In Sec.~IV, the concrete expressions of the former $h_\alpha(t)$  are presented for  individual chirping binaries. Assuming that the noise $n_\alpha(t)$ is statistically stationary, we  take  its Fourier transform as 
\beq
{\tilde n}_\alpha(f)=\int_{-\infty}^\infty dt e^{2\pi if t} n_\alpha(t),
\eeq
and define its spectrum $S_h(f)$ by
\beq
\lla  {\tilde n}_\alpha(f) {\tilde n}_\alpha(f')^*\rra=\frac12\delta(f-f') S_h(f).
\eeq
Here $\lla \cdot \rra$ represents an ensemble average. Since we only use the effectively L-shaped interferometers whose noise spectra are identical for given missions (DECIGO and BBO), we omitted the label $\alpha$ for  the spectrum.
In this paper we conservatively deal with an appropriate frequency regime $f\ge 0.2$Hz, considering the potential effects of  WD/WD binaries \cite{farmer}.

For its reference design parameters \footnote{We use the following parameters; the arm length 1000 km, the output laser  power 10 W with wavelength $\lambda=532$nm,
 the mirror diameter 1 m with its mass 100 kg, and 
the finesse of FP cavity  10 \cite{kawamura2006}.}, one can construct two L-shaped interferometers from a single triangular detector unit (see Fig.~\ref{ae}). Their noises are uncorrelated and the noise spectrum of such an L-shaped DECIGO is given by
\begin{equation}
S_h^{\mathrm{DECIGO}}(f)=7.05\times 10^{-48}\left[1+\left(\frac{f}{f_p}\right)^2\right] 
                         +4.8\times 10^{-51}\left(\frac{f}{1\ \mathrm{Hz}} \right)^{-4}\frac{1}{1+\left(\frac{f}{f_p}\right)^2} 
                         +5.33\times 10^{-52}\left(\frac{f}{1\ \mathrm{Hz}} \right)^{-4} \ \mathrm{Hz^{-1}}, \label{decigo-noise}
\end{equation}
%
with $f_p=7.36 \mathrm{Hz}$. 
The first, second and third  terms represent the shot noise, the radiation pressure noise and the acceleration noise, respectively. The above noise corresponds to rescaling the DECIGO noise spectrum in~\cite{kent2} by a factor $(\sqrt{3}/2)^{-2}$. This is due to the fact that in~\cite{kent2} the effect of the angle between detector arms being $60^\circ$ has been accounted for in the waveform amplitude while in this paper we include such an effect in the detector noise.

\begin{figure}[t]
  \includegraphics[scale=.35,clip]{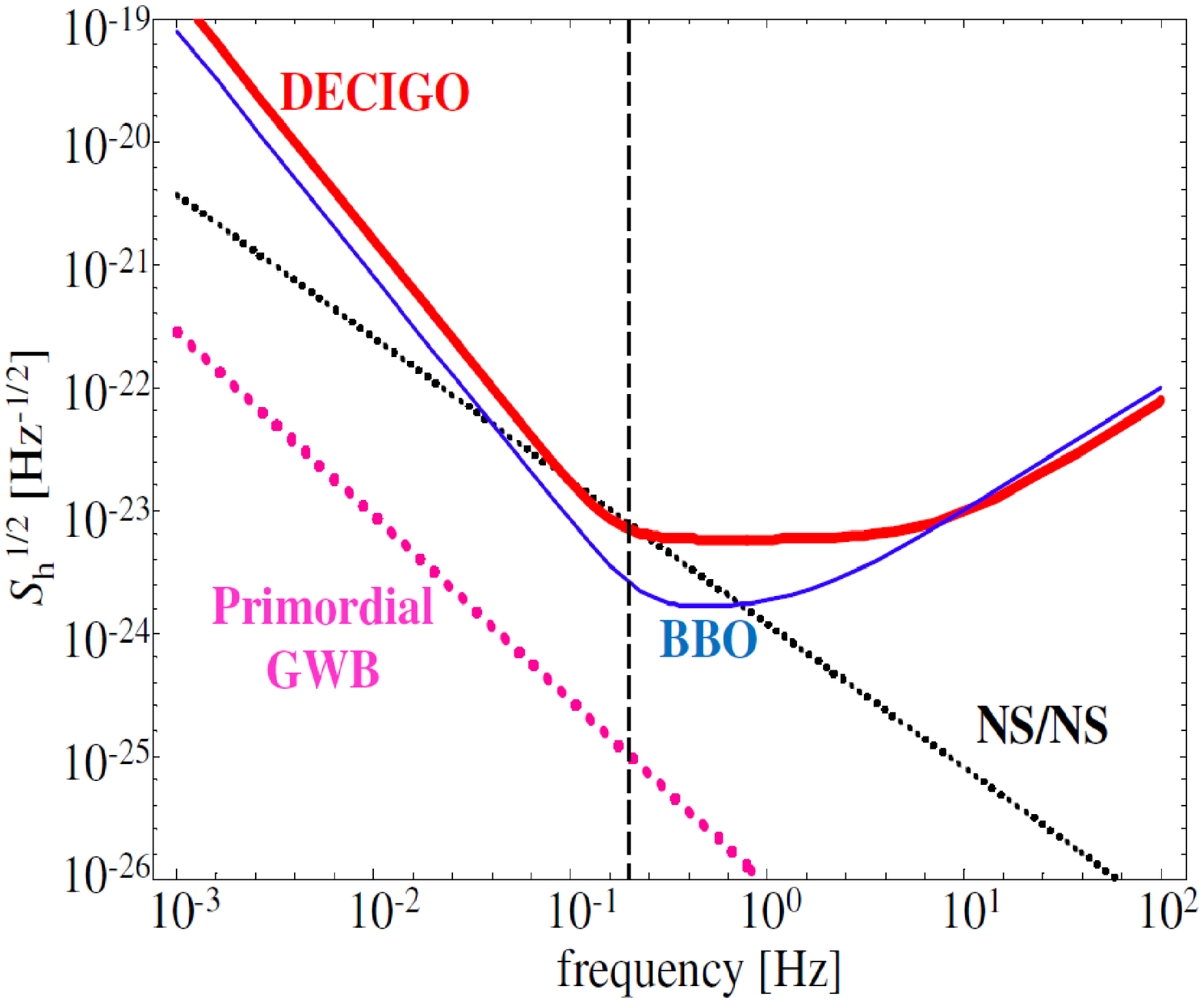} 
 \caption{\label{noise-decigo} The solid curves show the noise spectra  for DECIGO (thick red) and BBO  (thin blue). Note that these noise curves are the sky-averaged ones (a factor of $\sqrt{5}$ larger than the original ones). The (black) thin dotted line represents the expected total amplitude of NS/NS foreground and the (purple) thick dotted line represents the  primordial GW background corresponding to $\Omega_{\mathrm{GW}}=10^{-16}$. The (black) dashed line at $f=0.2$Hz indicates the upper frequency cutoff of WD/WD confusion noises.}
\end{figure}

For BBO with its reference parameters \footnote{The followings are the model parameters;  arm length 50000 km,  the  laser   power 300 W with wavelength $\lambda=500$nm, and the mirror diameter 3.5 m.},  we fit its noise curve shown in~\cite{cutlerharms} and use the following (non sky-averaged) expression;
\begin{equation}\label{bbon}
S_h^{\mathrm{BBO}}(f)=2.00\times 10^{-49} \left(\frac{f}{1\ \mathrm{Hz}} \right)^2
                         +4.58\times 10^{-49}
                         +1.26\times 10^{-51}\left(\frac{f}{1\ \mathrm{Hz}} \right)^{-4} \ \mathrm{Hz^{-1}} 
\end{equation}
that is almost {equivalent}  to the spectrum used in Cutler and Holz~\cite{cutlerholz} {(up to a factor 5 corresponding to sky-averaging)}.
In Fig~\ref{noise-decigo}, we present these noise spectra for each L-shaped interferometer.
 Here, in order to compare with the characteristic amplitude of a NS/NS binary, we {show} the sky averaged forms of the spectra which are a factor of 5 larger than the original ones expressed in Eqs.~(\ref{decigo-noise}) and~(\ref{bbon}). 
As shown in Fig.~\ref{noise-decigo}, the noise spectrum of DECIGO is $\sim 3$ times larger than that of BBO.
We mainly use the spectral shape of DECIGO for our numerical evaluation below and later we discuss its required level for sufficient NS/NS cleaning.
We also show that, in appropriately normalized forms,  our results for DECIGO are almost the same as those for BBO.  Therefore,  our analyses can be easily applied to BBO as well.


\section{\label{sec-GWB}GRAVITATIONAL WAVE BACKGROUND FROM NS/NS BINARIES}

Magnitude of a stochastic GW is characterized by its fractional  energy density $\Omega_{\mathrm{GW}}$ per logarithmic frequency interval as
\begin{equation}
\Omega_{\mathrm{GW}}\equiv\frac{1}{\rho_c}\frac{d \rho_{\mathrm{GW}}}{d \ln f}, \label{omega-gw}
\end{equation}
where $\rho_{\mathrm{GW}}$ is the energy density of the GWs and $\rho_c\equiv \frac{3H_0^2}{8\pi}$ is the critical energy density of the universe with the current Hubble parameter $H_0=70h_{70}$ km/s/Mpc with $h_{70}=1$ for our fiducial value~\cite{komatsu}.
For the total GW foreground by cosmological NS/NS binaries,  we apply the convenient formula given by Phinney~\cite{phinney} as
%
\begin{equation}
\Omega_{\mathrm{GW}}^{\mathrm{NS}}=\frac{8\pi^{5/3}}{9}\frac{1}{H_0^2}\mathcal{M}^{5/3} f^{2/3}\int _{0}^{\infty} dz \frac{\dot{n}(z)}{(1+z)^{4/3}H(z)}. \label{omega-NS}
\end{equation}
%
Here $\mathcal{M}$ is the chirp mass defined as $\mathcal{M}\equiv(m_1 m_2)^{3/5} (m_1+m_2)^{-1/5}$ with the two masses  $m_1$ and $m_2$ of binaries.  
In this paper, we assume $m_1=m_2=1.4\so$ ($\so=2.0\times 10^{33}$g) for NS/NS binaries.
The function 
$\dot{n}(z)$ is the NS/NS merger rate per proper time per comoving volume at redshift $z$, and the Hubble parameter $H(z)$ is given by 
\begin{equation}
H(z)\equiv H_0\sqrt{\Omega_m(1+z)^3+\Omega_{\Lambda}}
\end{equation}
for the  $\Lambda$CDM cosmology assumed in this paper  with the cosmological parameters  $\Omega_{m}=0.3$ and $\Omega_{\Lambda}=0.7$.  We re-express the rate
$\dot{n}(z)$   in terms of the current merger rate $\dot{n}_0$  and the redshift dependence $s(z)$  as $\dot{n}(z)=\dot{n}_0 \times s(z)$.
In this paper, we use  $\dot{n}_0=10^{-7}$ Mpc$^{-3}$ yr$^{-1}$ as a fiducial value.
For the redshift dependence $s(z)$, we adopt the following  piecewise linear fit ~\cite{cutlerharms, seto-lense}  based on Ref.~\cite{schneider};
\begin{eqnarray}
s(z)=\left\{ \begin{array}{ll}
1+2z & (z\leq 1) \\
\frac{3}{4}(5-z) & (1\leq z\leq 5) \\
0 & (z\geq 5). \\
\end{array} \right.
\end{eqnarray}
Then, the magnitude  $\Omega_{\mathrm{GW}}^{\mathrm{NS}}$ is numerically  evaluated as
\begin{equation}
\Omega_{\mathrm{GW}}^{\mathrm{NS}}(f)=1.73\times 10^{-12} h^{-3}_{70} \left( \frac{\mathcal{M}}{1.22 M_{\odot}} \right)^{5/3} 
                                                            \left( \frac{f}{1 \mathrm{Hz}}  \right)^{2/3} 
                                                             \left( \frac{\dot{n}_0}{10^{-7} \mathrm{Mpc}^{-3} \mathrm{yr}^{-1}}  \right). \label{nsb}
\end{equation}
The total (sky-averaged) GW foreground spectrum $S_h$ and the normalized energy density  $\Omega_{\mathrm{GW}}$ have the relation~\cite{cutlerharms} 
\begin{equation}
S_h=\frac{4}{\pi}f^{-3}\rho_c\Omega_{\mathrm{GW}}, \label{confusion}
\end{equation}
and, with Eqs.~(\ref{nsb}) and (\ref{confusion}), we obtain 
\begin{equation}
\sqrt{S_h^{\mathrm{NS}}}(f)=1.17\times 10^{-24}  h^{-1/2}_{70} \left( \frac{\mathcal{M}}{1.22 M_{\odot}} \right)^{5/6} \left( \frac{\dot{n}_0}{10^{-7} \mathrm{Mpc}^{-3} \mathrm{yr}^{-1}}  \right)^{1/2}
                                       \left( \frac{f}{1 \mathrm{Hz}}  \right)^{-7/6} \mathrm{Hz}^{-1/2}.
\end{equation}
For a PGWB,  we have
\begin{equation}
\sqrt{S_h^{\mathrm{PGWB}}}(f)=8.85\times 10^{-27} h_{70} \left( \frac{\Omega_{\mathrm{GW}}(f)}{10^{-16}} \right)^{1/2}
                                       \left( \frac{f}{1 \mathrm{Hz}}  \right)^{-3/2}\mathrm{Hz}^{-1/2}.
\end{equation}
These relations are also shown in Fig.~\ref{noise-decigo}.
As mentioned in the introduction, to detect a PGWB as small as $\Omega_{\mathrm{GW}}\sim 10^{-16}$ around $f\sim 1$Hz, we need to clean the NS/NS foreground and reduce its residual by 5 orders of magnitude in terms of $\Omega_{\mathrm{GW}}$.

\section{\label{sec-det} GW from binary AND DETECTOR RESPONSE}
In order to subtract the GW foreground formed by cosmological NS/NS binaries,  it is essential to individually identify them in the data streams of detectors.  Here, we summarize the basic formulae required for evaluating the SNRs of these binaries.

\begin{figure}[t]
  \includegraphics[scale=.5,clip]{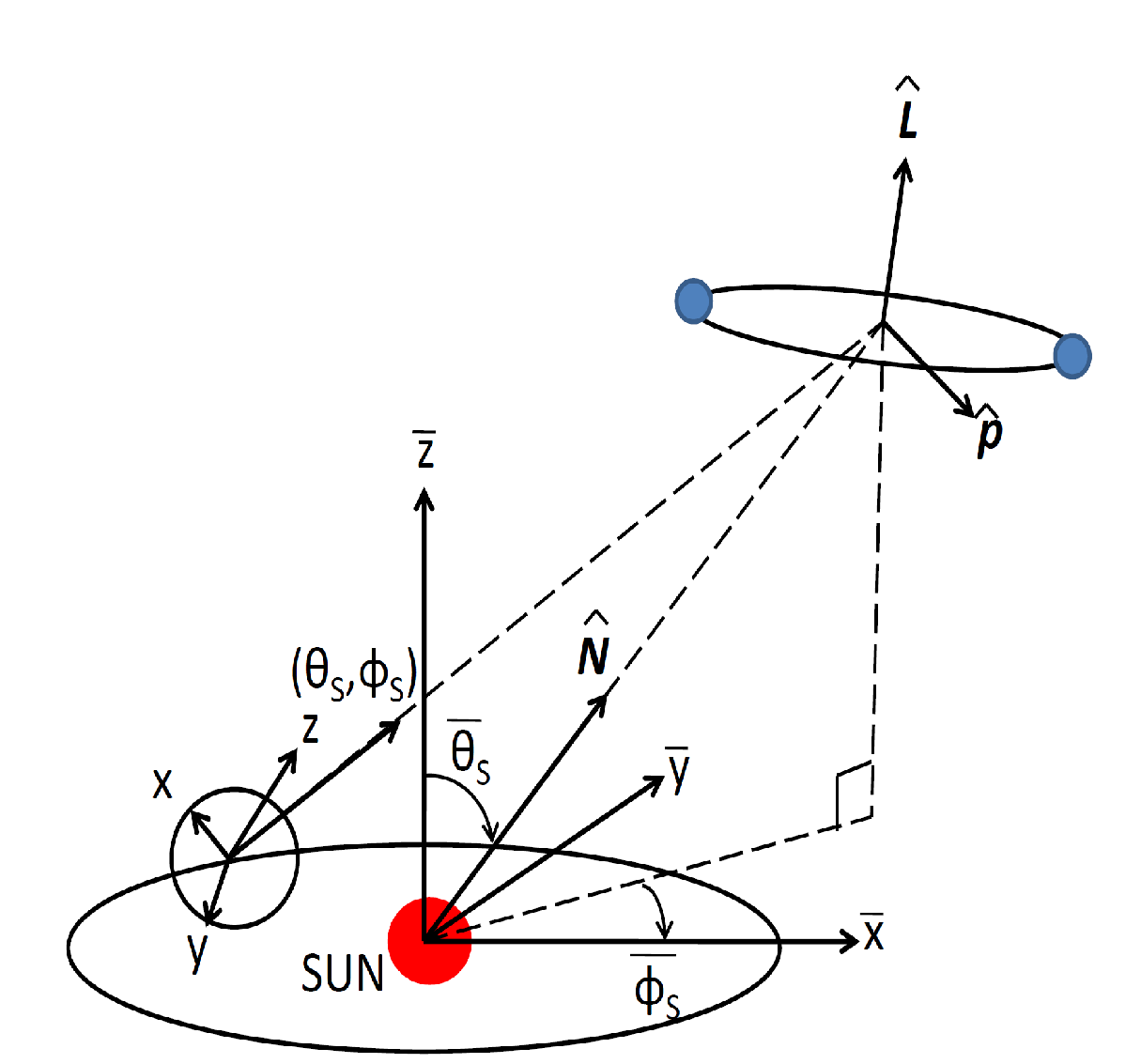} 
 \caption{\label{fig1} We use two types of coordinates: (i) a barycentric frame $(\bar{x},\bar{y},\bar{z})$ tied to the ecliptic and centered in the solar system barycenter, (ii) an detector frame $(x,y,z)$  attached to the detector as in Fig.~\ref{ae}. }
\end{figure}

 First, we provide expressions for the GWs coming from a NS/NS binary.
Since we start our calculation from 0.2 Hz where  eccentricity of the binary is expected to be small, we can reasonably  assume a circular orbit for studying detectability of the binary.
From the quadrupole formula of GWs, the waveforms of + and $\times$ modes in the principal polarization coordinate are given by
\begin{eqnarray}
h_{+}(t)&=&A_{+}\cos\phi(t), \\
h_{\times}(t)&=&A_{\times}\sin\phi(t),
\end{eqnarray} 
with the phase function  $\phi(t)$ and  the amplitudes
\begin{eqnarray}
A_{+}&=&\frac{2m_1m_2}{rD_L}(1+(\hat{\bm{L}}\cdot\hat{\bm{N}})^2), \label{aplus} \\
A_{\times}&=&-\frac{4m_1m_2}{rD_L}(\hat{\bm{L}}\cdot\hat{\bm{N}}). \label{across}
\end{eqnarray}
Here $r$ is the orbital separation of the binary, $D_L$ is the luminosity distance to the source, $\hat{\bm{L}}$ is the unit vector parallel to the orbital angular momentum and $\hat{\bm{N}}$ is the unit vector pointing towards the source from the observer (see Fig.~\ref{fig1}).

Next, we deal with  responses of GW detectors.
Following Ref.~\cite{cutler1998}, we introduce two Cartesian reference frames: (i) a barycentric frame $(\bar{x},\bar{y},\bar{z})$ tied to the ecliptic and centered in the solar system barycenter, with $\hat{\bar{\bm{z}}}$ (unit vector in ${\bar{\bm{z}}}$ direction) normal to the ecliptic, (ii) an detector frame $(x,y,z)$ attached to the detector, with the direction  $\hat{\bm{z}}$ normal to the detector plane (see Figs.~\ref{ae} and \ref{fig1}). 

While each unit of DECIGO or BBO changes its position and orientation as a function of time, we first discuss its response by fixing the geometry of the source-detector system.  As already discussed in Sec.~{\ref{sec-noise}}, the triangle unit can be effectively regarded as the two L-shaped interferometers I and II. 
Each output of these interferometers is written as
%
%
\begin{equation}
h_{\mathrm{\alpha}}(t)= A_{+} F_{\mathrm{\alpha}}^+(\theta_{\mathrm{S}},\phi_{\mathrm{S}},\psi_{\mathrm{S}}) \cos\phi(t)
          + A_{\times} F_{\mathrm{\alpha}}^{\times}(\theta_{\mathrm{S}},\phi_{\mathrm{S}},\psi_{\mathrm{S}}) \sin\phi(t) \label{output2}
\end{equation}
with the label for interferometers $\alpha$=I, II.
 The  beam pattern functions $F_{\mathrm{I}}^{+,\times}$  for the interferometer I are given by
%
\begin{eqnarray}
F_{\mathrm{I}}^{+}(\theta_{\mathrm{S}},\phi_{\mathrm{S}},\psi_{\mathrm{S}}) 
                &=&\frac{1}{2}(1+\cos^2 \theta_{\mathrm{S}}) \cos(2\phi_{\mathrm{S}}) \cos (2\psi_{\mathrm{S}})
                  -\cos(\theta_{\mathrm{S}}) \sin(2\phi_{\mathrm{S}}) \sin(2\psi_{\mathrm{S}}),  \\
F_{\mathrm{I}}^{\times}(\theta_{\mathrm{S}},\phi_{\mathrm{S}},\psi_{\mathrm{S}})
                &=&\frac{1}{2}(1+\cos^2 \theta_{\mathrm{S}}) \cos(2\phi_{\mathrm{S}}) \sin (2\psi_{\mathrm{S}})
                  +\cos(\theta_{\mathrm{S}}) \sin(2\phi_{\mathrm{S}}) \cos(2\psi_{\mathrm{S}}), \label{beam-patternI}
\end{eqnarray} 
and those for the interferometer II are expressed as
\begin{eqnarray}
F_{\mathrm{II}}^{+}(\theta_{\mathrm{S}},\phi_{\mathrm{S}},\psi_{\mathrm{S}})&=&F_{\mathrm{I}}^{+}(\theta_{\mathrm{S}},\phi_{\mathrm{S}}-\pi/4,\psi_{\mathrm{S}}), \\
F_{\mathrm{II}}^{\times}(\theta_{\mathrm{S}},\phi_{\mathrm{S}},\psi_{\mathrm{S}})&=&F_{\mathrm{I}}^{\times}(\theta_{\mathrm{S}},\phi_{\mathrm{S}}-\pi/4,\psi_{\mathrm{S}}). \label{beam-patternII}
\end{eqnarray}
Here, the two angles $(\theta_{\mathrm{S}},\phi_{\mathrm{S}})$ represent the direction of the source in the detector frame and $\psi_{\mathrm{S}}$ is the polarization angle given as
\begin{equation}
\tan\psi_{\mathrm{S}}=\frac{\hat{\bm{L}}\cdot\hat{\bm{z}}-(\hat{\bm{L}}\cdot\hat{\bm{N}})(\hat{\bm{z}}\cdot\hat{\bm{N}})}
                                  {\hat{\bm{N}}\cdot(\hat{\bm{L}}\times\hat{\bm{z}})}. \label{tanpsi}
\end{equation}
This angle characterizes the orientation of the vector $\hat{\bm{L}}$ projected on the transverse plane (see Fig. 1 of Ref.~\cite{apostolatos}).

%
%

By using the stationary phase approximation, we obtain the Fourier transform of the signal as
\begin{equation}
\tilde{h}_\alpha(f)=\mathcal{A}f^{-7/6}e^{i\Psi (f)} \left[ \frac{5}{4}A_{\mathrm{pol},\alpha} \right] e^{-i \left( \varphi_{\mathrm{pol},\alpha}+\varphi_D \right)}, \label{waveform}
\end{equation}
where the amplitude $\mathcal{A}$ is given by the redshifted chirp mass  $\mathcal{M}_z\equiv (1+z)\mathcal{M}$ as 
\begin{equation}
\mathcal{A}=\frac{1}{\sqrt{30}\pi^{2/3}}\frac{\mathcal{M}_z^{5/6}}{D_L}.  \label{amp-noangle}
\end{equation}
The polarization amplitude $A_{\mathrm{pol},\alpha}$ is defined by
%
\begin{equation}
A_{\mathrm{pol},\alpha}=\sqrt{(1+(\hat{\bm{L}}\cdot\hat{\bm{N}})^2)^2(F_{\alpha}^{+})^2+4(\hat{\bm{L}}\cdot\hat{\bm{N}})^2(F_{\alpha}^{\times})^2}. \label{Apol} 
\end{equation}
%

In Eq.~(\ref{waveform}), $\Psi(f)$ is the phase in frequency domain, $\varphi_{\mathrm{pol},\alpha}$ is the polarization phase, and $\varphi_{D}$ is the Doppler phase which denotes the difference between the phase of the wavefront at the detector and the one at the solar system barycenter.
Since we are only interested in the signal to noise ratio (SNR) of binary GWs, we do not need explicit forms for the phases.

Next we define the total SNR  for one triangle unit  by 
\begin{equation}
SNR^2=4\sum_{\alpha=\mathrm{I},\mathrm{II}}\int^{f_{\mathrm{max}}}_{f_{\mathrm{min}}} \frac{|\tilde{h}(f)|^2}{S_h(f)} df, \label{rho2}
\end{equation}
and it has the following relation with respect to the geometrical parameters
\begin{equation}
\begin{split}
SNR^2 \propto & \left( \frac{1+\cos^2 i}{2} \right)^2 \left[ \left( \frac{1+\cos^2 \theta_{\mathrm{S}}}{2} \right)^2 \cos^2 (2\psi_{\mathrm{S}})+\cos^2\theta_{\mathrm{S}} \sin^2(2\psi_{\mathrm{S}}) \right] \\
                     & +\cos^2 i \left[ \left( \frac{1+\cos^2 \theta_{\mathrm{S}}}{2} \right)^2 \sin^2 (2\psi_{\mathrm{S}})+\cos^2\theta_{\mathrm{S}} \cos^2(2\psi_{\mathrm{S}}) \right], \label{rho2-angle}
\end{split}
\end{equation}
where $i$ is the inclination angle defined as $\cos i \equiv \hat{\bm{L}}\cdot\hat{\bm{N}}$.  It is important to note that the right-hand side of this relation does not depend on the azimuthal angle $\phi_S$.  This is related to the orthonormality of the two effective detectors I and II.

\begin{figure}[tbp]
  \includegraphics[scale=.5,clip]{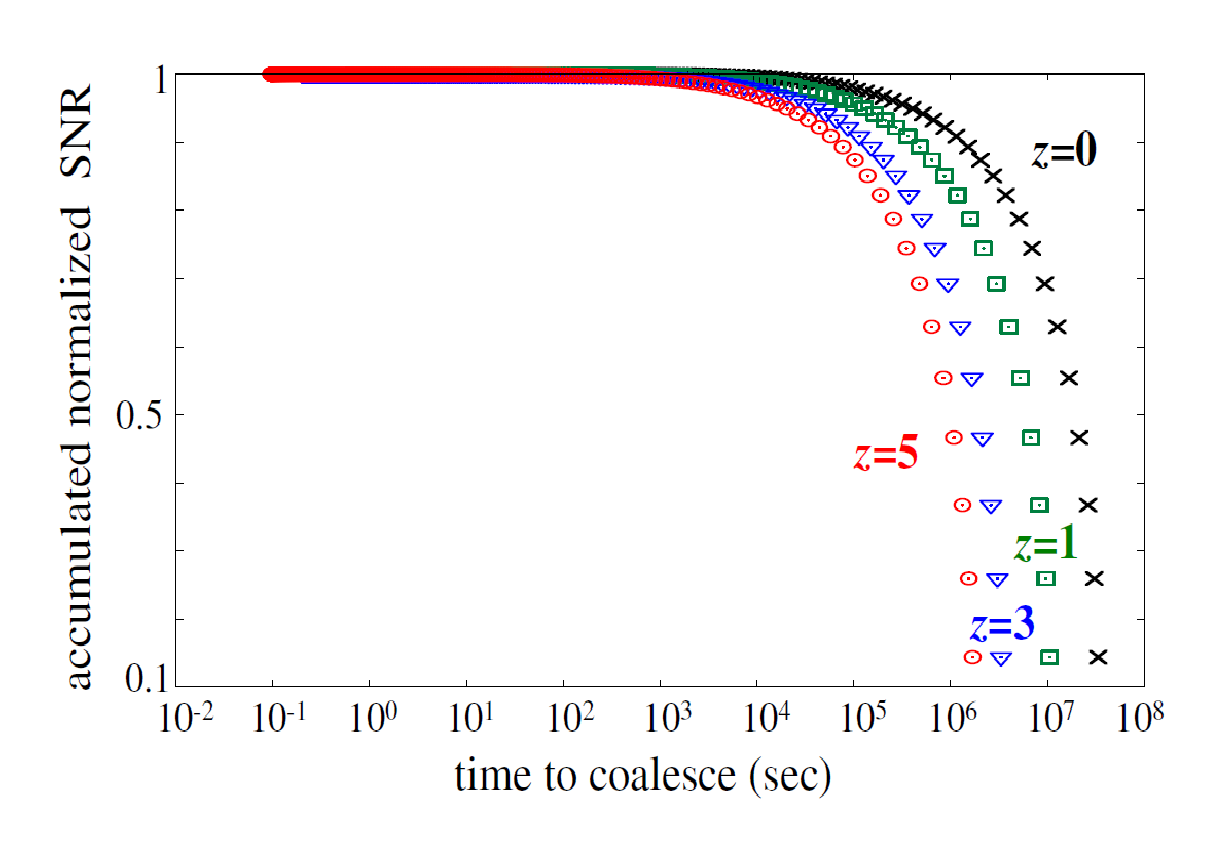} 
 \caption{\label{t-sn} This figure shows the accumulated averaged SNR of each (1.4+1.4)$\so$ NS/NS binary against time to coalesce.
The (red) circular plots, the (blue) triangular plots, the (green) square plots and the (black) crosses correspond to the one with the source redshift $z=5$,3,1 and 0, respectively.
These values have been normalized so that the accumulated SNRs become 1 at the time corresponding to $f$=100Hz.}
\end{figure}

Up to 2PN order,  we obtain the relation  between the GW frequency $f$ and the time $t$ as~\cite{berti}
\begin{equation}
\begin{split}
t(f)=t_c-\frac{5}{256}\mathcal{M}_z(\pi \mathcal{M}_zf)^{-8/3} \biggl[1&+\frac{4}{3}\left( \frac{743}{336}+\frac{11}{4}\eta \right) x -\frac{32}{5}\pi x^{3/2}  \\
       & +2\left( \frac{3058673}{1016064}+\frac{5429}{1008}\eta+\frac{617}{144}\eta^2 \right) x^2 \biggr]. \label{tf}
\end{split}
\end{equation}
Here $t_c$ is the coalescence time, $\eta=m_1m_2/(m_1+m_2)^2$ is the reduced mass ratio, and $x$ is  defined as  $x\equiv (\pi (1+z)M f)^{2/3}$ with the total mass $M=m_1+m_2$.
Since we neglect NS spins in this paper, we have omitted the spin-orbit coupling and the spin-spin coupling.
When we only take the leading part and set $t_c=0$,  the time before the coalescence is given by 
\begin{equation}
-t(f)=1.04\times 10^{7} \left( \frac{1+z}{2} \right)^{-5/3} \left( \frac{\mathcal{M}}{1.22 M_{\odot}} \right)^{-5/3} \left( \frac{f}{0.2 \mathrm{Hz}} \right)^{-8/3} \mathrm{sec}.
\end{equation}
For $z=0,3 $ and $5$, the time $-t(f=0.2 \mathrm{Hz})$ becomes $3.32\times 10^7 \mathrm{sec}$, $3.29\times 10^6 \mathrm{sec}$ and  $1.67\times 10^6 \mathrm{sec}$, respectively.
In Fig.~\ref{t-sn}, we plot the accumulated averaged SNRs (normalized appropriately) against the time before coalescence for NS/NS binaries at $z=0,1,3$ and 5.
Since there would be WD/WD binary confusion noises below $f=0.2$Hz (see Fig.~\ref{noise-decigo}), we performed calculations from $f=0.2$Hz to $f=100$Hz.
From this figure, it can be seen that the effective observation time depends strongly on the source redshift.
There is almost 1 yr observation time for a binary at $z=0$, but it is less than a month for a binary at $z=5$.

In reality, each unit of  DECIGO or BBO is moving around the sun with the period $T=1$yr and characterized by the position angles
\begin{equation}
\bar{\theta}(t)=\pi/2, \qquad \bar{\phi}(t)=2\pi t/T+c_0
\end{equation}
with a constant $c_0$ (in this paper we take $c_0=0$).
As we described in the previous paragraph, the observation times become longer for lower redshift binaries and it is important to take the effect of detector motions into account to perform more practical analyses.
The time dependent amplitude $A_{\mathrm{pol,\alpha}}$ essentially corresponds to the right hand side of Eq.~(\ref{rho2-angle}) which is a function of $\theta_{\mathrm{s}}$, $\psi_{\mathrm{s}}$ and $i$ in the moving detector frame.
We can change these variables to the angles $(\bar{\theta}_{\mathrm{S}},\bar{\phi}_{\mathrm{S}},\bar{\theta}_{\mathrm{L}}, \bar{\phi}_{\mathrm{L}})$ in the fixed frame by using the formulae shown in Appendix.



Next we summarize the angular averaged SNR  for binaries at a redshift $z$ observed by a network of totally $N_{\mathrm{unit}}$ units of detectors, assuming that each contains two independent interferometers.  Using the simple identity $\lla A_{\mathrm{pol},\alpha}^2\rra=(4/5)^2$ for the angular average, we have 
\begin{equation}
\lla SNR_z^2\rra=8N_{\mathrm{unit}} \mathcal{A}^2 \int^{f_{\mathrm{max}}}_{f_{\mathrm{min}}} \frac{f^{-7/3}}{S_h(f)} df \label{rhoz}.
\end{equation}
For a given position, the SNR of a binary becomes minimum for an edge-on geometry, and we have  $\lla A_{\mathrm{pol},\alpha}^2\rra_{\mathrm{edge-on}}=1/5$ for this subclass.  Then the averaged SNR for these binaries is given as\beq
\bar{\rho}_z^2\equiv \lla  SNR_z^2 \rra_{\mathrm{edge-on}}=\frac5{16} \lla  SNR_z^2 \rra, \label{edgem}
\eeq
and is independent on the network geometry.
Hereafter we mainly characterize the total sensitivity of a detector network by the parameter $\bar{\rho}_5$ defined at the most distant redshift  $z=5$ in our fiducial model.  This would be a convenient measure for discussing a network sensitivity required to identify foreground binaries.  
With $N_{\mathrm{unit}}=4$, $f_{\mathrm{min}}=0.2$Hz and $f_{\mathrm{max}}=100$Hz,  we have the network sensitivity  $\bar{\rho}_5=10.4$ for DECIGO and $\bar{\rho}_5=33.6$ for BBO.
The result
$\bar{\rho}_5$ for BBO here is $\sim 20\%$ smaller than that in Cutler \& Harms \cite{cutlerharms}, due to the difference in the lower frequency cut-off $f_{\mathrm{min}}$.

 Since Eq.~(\ref{edgem}) is a mean value, a given network must satisfy a constraint
\beq
G_z \equiv\frac{SNR_{z,\mathrm{min}}}{\bar{\rho}_z}\le 1 \label{defg}
\eeq
with respect to its minimum value $SNR_{z,\mathrm{min}}$ for binaries at a  redshift $z$.  The equality holds only when the SNR is independent on the direction and the polarization angles, {\it e.g.} for infinite number of units with random orientations (see also \cite{Boyle:2010gc}).
{Note that the ratio  $G_z$  does not depend on the overall sensitivity of the detector (but on the shape of its noise spectrum).}

\section{\label{sec-dead}Analytic study for the minimum SNR}
Our primary interest in this paper is whether DECIGO or BBO can successfully identify the cosmological NS/NS binaries in performing their subtraction with sufficient accuracy to uncover an underlying stochastic background.  The critical aspect here is the profile of the lower end of the probability distribution function of their  SNRs. While we extensively perform numerical analyses in the next section, the simple analytical studies in this section would help us in understanding basic geometrical relations and also to interpret our numerical results. 
In this section,  we fix the redshift $z$ of  binaries and do not include the annual motions of detectors for simplicity.  The latter prescription would be a reasonable approximation for binaries at a large redshift $z$.

\begin{figure}[tbp]
  \includegraphics[scale=.6,clip]{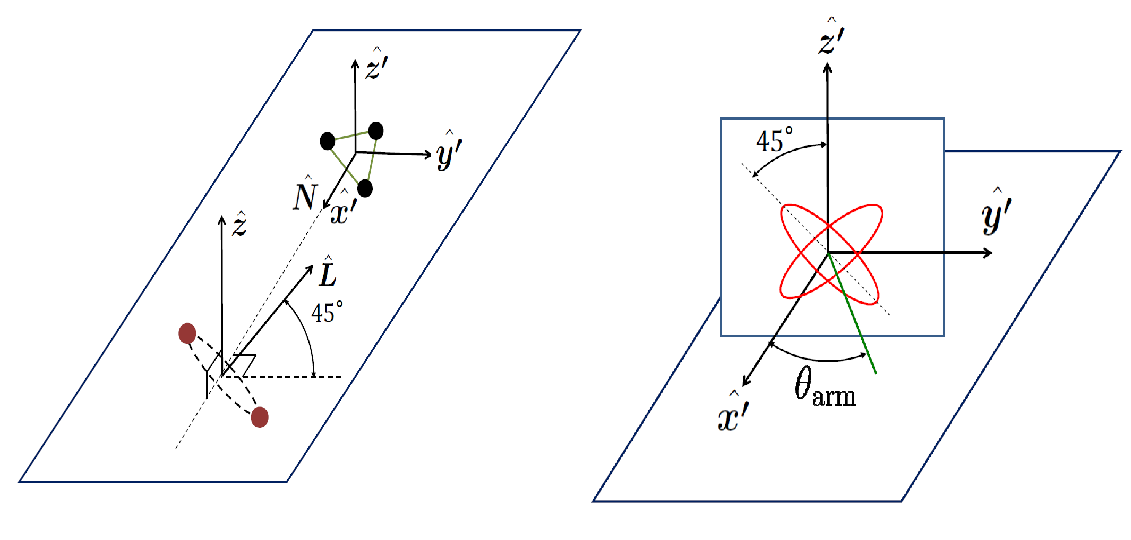} 
 \caption{\label{dead-angle} (Left) The dead angle for gravitational wave interferometers with one detector plane. $\hat{\bm{N}}$ is the unit vector from the detector to the binary, the unit vector $\hat{\bm{z}}$ is normal to the detector plane and the unit vector $\hat{\bm{L}}$ represents the orientation of the orbital angular momentum of the binary. $\hat{\bm{L}}$ is perpendicular to $\hat{\bm{N}}$ and has an angle 45$^{\circ}$ from $\hat{\bm{z}}$. We also introduce a new Cartesian  coordinate $(x',y',z')$ at the detector  with $\hat{\bm{x}}'=\hat{\bm{N}}$ and  $\hat{\bm{z}}'=\hat{\bm{z}}$.
(Right) The linear polarization pattern on the $y'-z'$ plane for the GW signal from a dead-angled binary (note that the other orthogonal  polarization mode vanishes for the edge-on binary). An arbitrary detector arm in the $x'-y'$ plane is completely blind to this signal.}
\end{figure}

\begin{figure}[tbp]
  \includegraphics[scale=.4,clip]{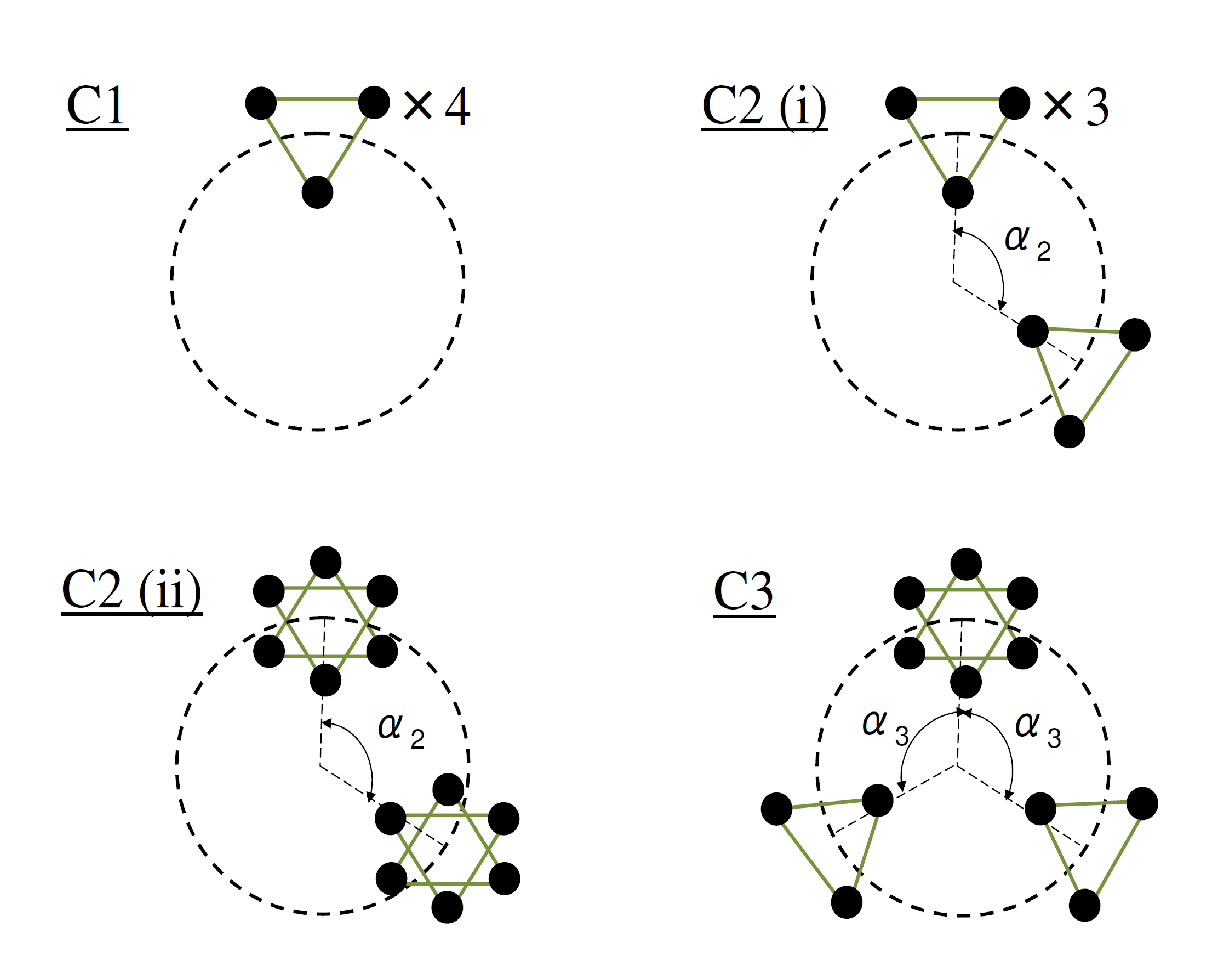} 
 \caption{\label{config-cases} Configurations made by four units of detectors.
In reality, each triangle unit is  inclined by $60^\circ$ from the orbital plane.
(Top Left) We put all units on one site. There exists only one detector plane for this configuration. We label this as configuration ``C1". 
(Top Right) We place three units on one site and put the remaining one with opening angle $\alpha_2$.
There are 2 detector planes for this configuration ``C2(i)".
(Bottom Left)  We place two out of four units on one site and the remaining two units on another site with  a separation angle of $\alpha_2$.
We label this as configuration ``C2(ii)".
(Bottom Right) {This configuration ``C3" is the same as the one shown in Fig~\ref{default}}. There exist three detector planes for this configuration.}
\end{figure}

Under these conditions, we discuss the minimum value  $SNR_{z,\mathrm{min}}$ mostly dealing with edge-on binaries. As we see below, we have $ SNR_{z,\mathrm{min}}=0$ with one detector plane, due to the apparent dead angle described in Fig.~\ref{dead-angle}. With multiple detector planes, the dead angle generally disappears,  resulting in $SNR_{z,\mathrm{min}}>0$.  Thus we evaluate the performance of a network with multiple planes, using the normalized measure $G$ defined in Eq.~(\ref{defg}) (in this section, we simply denote $G_z$ as $G$ since it does not depend on $z$ for the static detector networks).  The basic question here is how far we can increase the value $G$ with a small number of units. As concrete examples, we pick up the specific configurations shown in Fig.~\ref{config-cases}.

\subsection{one detector plane}

As a first step, let us consider the situation only with a single detector plane (C1 in Fig.~\ref{config-cases}). From Eq.~(\ref{rho2-angle}), we can straightforwardly confirm that the total SNR becomes identical to zero,  when the following 3 conditions are simultaneously satisfied;  $\cos i=0$, $\theta_{\mathrm{S}}=\pi/2$ and $\cos 2\psi_{\mathrm{S}}=0$.
This configuration is illustrated in the left panel of Fig.~\ref{dead-angle}. 
Here, we introduce a new Cartesian coordinate $(x',y',z')$ with $\hat{\bm{x'}}=\hat{\bm{N}}$ and  $\hat{\bm{z'}}=\hat{\bm{z}}$. 
For an edge-on binary, there is only one linear  polarization mode (see Eqs.~(\ref{waveform}) and~(\ref{Apol}) ).
Therefore we focus on the relevant  mode shown in the right panel of Fig.~\ref{dead-angle}.
On the $y'-z'$ plane  its principal axis is tilted by  $45^{\circ}$ from the normal vector $\hat{\bm{z'}}$. 
We can evaluate the response of a detector arm with an angle $\theta_\mathrm{arm}$ from the direction $\hat{\bm{x'}}$ as~\cite{cutler1998}
\beq
\begin{pmatrix}
\cos\theta_\mathrm{arm}, & \sin\theta_\mathrm{arm}, & 0
\end{pmatrix}
\begin{pmatrix}
0 & 0 & 0 \\
0 & 0 & 1 \\
0 & 1 & 0
\end{pmatrix}
\begin{pmatrix}
\cos\theta_\mathrm{arm} \\
\sin\theta_\mathrm{arm} \\
0
\end{pmatrix}
=0.
\eeq
Since this equation holds for an arbitrary $\theta_\mathrm{arm}$,  any GW detector on this plane is completely blind to this incoming signal. 

By expanding Eq.~(\ref{rho2-angle}) around the dead angles $(\theta_{\mathrm{S}},\psi_{\mathrm{S}},i)=(\pi/2, \pi/4, \pi/2)$, we get, up to the leading order,
\begin{equation}
\rho^2 \propto \frac{d\theta_{\mathrm{S}}^2}{4}+\frac{d\psi_{\mathrm{S}}^2}{4}+\frac{di^2}{4} \label{dead}
\end{equation}
with $(d\theta_{\mathrm{S}},d\psi_{\mathrm{S}},di)\equiv (\theta_{\mathrm{S}}-\pi/2, \psi_{\mathrm{S}}-\pi/4, i-\pi/2)$.  We will revisit this simple model later in Sec.~\ref{sec-num}.

\begin{figure}[tbp]
  \includegraphics[scale=.8,clip]{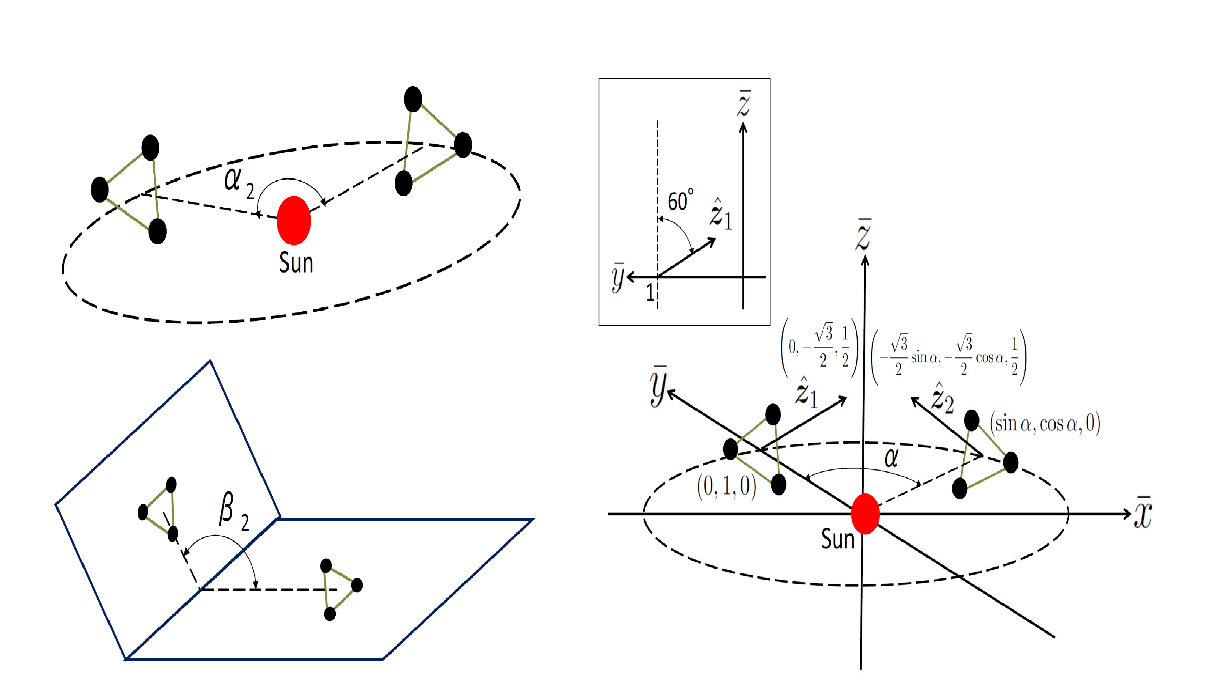} 
 \caption{\label{det-plane-angle} (Left) $\alpha_2$ denotes the separation angle between two detectors and $\beta_2$ represents the angle formed by the planes on which these two detectors lie. (Right) Positions of two detectors and unit normal vectors $\hat{\bm{z}}_1$ and $\hat{\bm{z}}_2$ are shown. Small panel on top left shows the $\bar{y}-\bar{z}$ plane, indicating that each detector tilts 60$^{\circ}$ inwards from $\bar{z}$ axis.  From the product $\hat{\bm{z}}_1\cdot\hat{\bm{z}}_2$, we have the correspondence $\cos\beta_2=(3\cos\alpha_2+1)/4$.}
\end{figure}

\subsection{two detector planes}

Next, we consider networks with two detector planes, namely configurations C2(i) and C2(ii)  shown in Fig.~\ref{config-cases}. 
In the present geometrical analysis, the network characterized by the position angle $\alpha_2$ in the top left panel of Fig.~\ref{det-plane-angle} is equivalent to the network defined by the opening  angle $\beta_2$ given in the bottom left panel of the same figure.  
These two angles are related with
\beq
\cos\beta_2=\frac{3\cos\alpha_2+1}4,
\eeq
which is derived by considering the inner product of the unit normal vectors $\hat{\bm{z}}_1$ and $\hat{\bm{z}}_2$ for each detector plane shown in the right panel of Fig.~\ref{det-plane-angle}.

\begin{figure}[tbp]
  \includegraphics[scale=.8,clip]{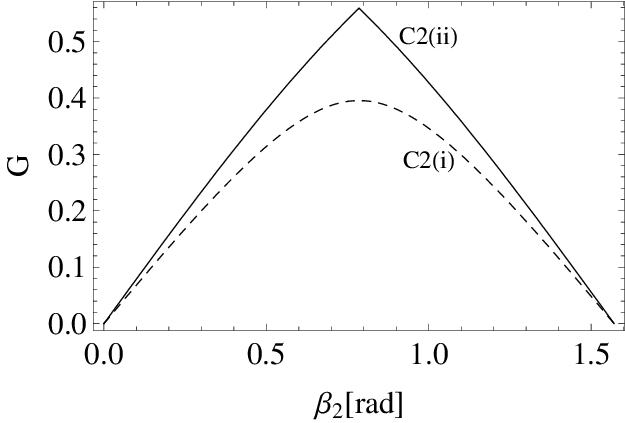} 
 \caption{\label{opt2d} The normalized ratio $G\equiv {SNR_{z,\mathrm{min}}}/{\bar{\rho}_z} $ for C2(i) and C2(ii) as  functions of the opening angle $\beta_2$. The optimal angle is $\beta_2=\pi/4$ (corresponding to $\alpha_2=52.4^\circ$).}
\end{figure}

With the dead angles for the previous subsection, it can be easily understood that we have $G=0$ only for the specific angles $\beta_2=0~({\rm mod}~ \pi/2)$.  For other angles, we find that the minimum value $SNR_{z,{\mathrm{min}}}(>0)$ is given by an edge-on binary at the direction of the intersection of two planes. {From Eq.~(\ref{rho2-angle})} we can show that, as a function of $\beta_2$, the maximum value of $G$ is realized at the optimal angle $\beta_2=\pi/4$ with 
\beq
G(\beta_2=\pi/4)=\frac{\sqrt{5r}}{2\sqrt{2}}.
\eeq
  Here the parameter $r (\le1/2)$ represents the relative weight {of the number of detector units between two sites} ({\it e.g.} $r=1/4$ for C2(i) and 1/2 for C2(ii)).   In Fig.~\ref{opt2d} we show the ratio $G$ for the configurations C2(i) and C2(ii).  The optimal angle $\beta_2=\pi/4$ corresponds to $\alpha_2=0.915$ ($52.4^\circ$), and we have $G(\pi/4)=\sqrt{5}/4\sqrt{2}=0.395$ for C2(i) and $\sqrt{5}/4=0.559$ for C2(ii).  For the latter  C2(ii), we can further derive a simple expression as
\beq
G(\beta_2)=\frac{\sqrt{5}}{2\sqrt{2}}\min[|\cos\beta_2|,|\sin\beta_2|].
\eeq
As one can see, when $\alpha_2$ equals to $\pi$ (maximum separation), the opening angle $\beta_2$ becomes $2\pi/3$ and this configuration is not the optimal choice.

Although Fig.~\ref{opt2d} is symmetric with respect to the value $\beta_2=\pi/4$, this does not mean that the situations are the same for $\beta_2 < \pi/4$ and $\beta_2 > \pi/4$.
For example, let us focus on two configurations with opening angles of $\beta_2=0$ and $\pi/2$.
$\beta_2=0$ corresponds to a configuration with only one detector plane while there are two detector planes for $\beta_2=\pi/2$.
Since each has some binaries that are completely insensitive to each configuration, they both give $G=0$.
However, the fraction of these dead-angle binaries to whole binaries is smaller for $\beta_2=\pi/2$ than the one for $\beta_2=0$.
Therefore we expect that the symmetry of Fig.~\ref{opt2d} is broken when we take the motions of detectors into account.

\subsection{three detector planes}
\label{three}

\begin{figure}[tbp]
  \includegraphics[scale=.8,clip]{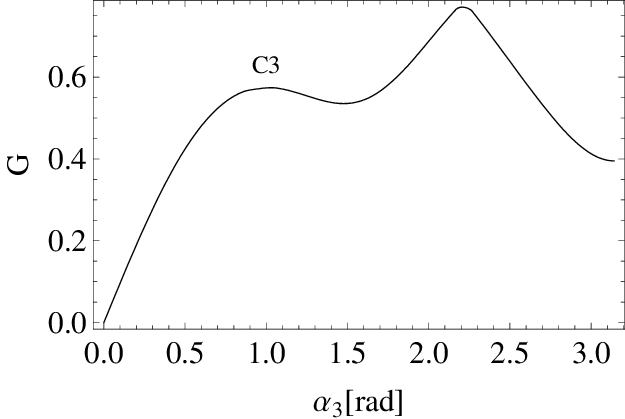} 
 \caption{\label{opt3d} The normalized ratio $G\equiv {SNR_{z,\mathrm{min}}}/{\bar{\rho}_z} $ for C3 {as a function of the detector separation angle $\alpha_3$} .  We have $G=0.771$ at the optimal angle $\alpha_3=2.205$ ($126.36^\circ$).}
\end{figure}

As an extension of the default configuration sketched in  Fig.~\ref{default}, we
relax the constraint $\alpha_3=120^\circ$
for the  angle between the three positions of units, and evaluate the
normalized ratio $G$ as a function of
$\alpha_3$. Our result is presented in Fig.~\ref{opt3d}.  The ratio $G$
takes its maximum value 0.771 at the
optimal angle $\alpha_3=2.205$ ($126.36^\circ$) that is slightly
larger than the default angle $\alpha_3=120^
\circ$.

For comparison, we examine a similar network but with evenly
weighted sensitivity (namely 1:1:1 instead of
2:1:1 as in Fig.~\ref{default}). In this case, the maximum of the ratio $G$ becomes
0.843 at the symmetric configuration
$\alpha_3=120^\circ$.  It seems reasonable that, as we increase the
weight of the overlapped units shown in Fig.~\ref{default},
the optimal angle $\alpha_3$ shifts to a larger value from the
symmetric value $\alpha_3=120^\circ$.  We also
evaluate the parameter $G$ when more than three detectors are
available on the ecliptic plane and evenly
weighted. In the large number limit, the asymptotic value can be given
analytically as $G=\sqrt{205}/16=0.895$~\cite{seto-annual}.
Note that even if we include the detector motions, this value is
unchanged and thus can be regarded as an useful
upper bound for the BBO/DECIGO-type orbital configurations.
We expect that as we increase the symmetry of a detector configuration, the effect of detector motion is suppressed. 

One might be interested in a  network formed by evenly weighted three
units with each normal to the three
orthogonal directions $(1,0,0)$, $(0,1,0)$ and $(0,0,1)$.  Even though
we do not have a direct orbital model for
space detectors, it could be an efficient network in terms of the
ratio $G$. Indeed, we have $G=5/\sqrt{30}=
0.913$ for this hypothetical network (see also \cite{Boyle:2010gc}).

In summary, using four equivalent units shown in Fig.~\ref{default} with a free
parameter $\alpha_3$, we can realize $G=0.771
$ that is at least 39\% larger than a network with two detector
planes.  The optimal angle $\alpha_3$ is $126.36
^\circ$ and somewhat larger than the default value $120^\circ$.
Here, we should remind that our
results in this section are derived by neglecting motions of
detectors, and valid for the binary sources at the high redshift
limit.  In the next section, we numerically
evaluate the parameter $G$ for realistic
cosmological binaries. These binaries have finite duration times in the
BBO/DECIGO band (see Fig.~\ref{t-sn}) and we should
include the detector motions when studying  their SNRs.

\section{NUMERICAL CALCULATIONS AND RESULTS  }
\label{sec-num}

\subsection{Set Up}

\quad In order to estimate how well we can subtract  the foreground GWs by NS/NS binaries, we performed  Monte Carlo simulations in the following manner.
First we uniformly divide redshift range  $z=[0,5]$ into 200 bins.
For each redshift bin, we randomly generate 10$^4$ sets of binary directions $(\bar{\theta}_{\mathrm{S}},\bar{\phi}_{\mathrm{S}})$ and orientations $(\bar{\theta}_{\mathrm{L}},\bar{\phi}_{\mathrm{L}})$, with $\cos\bar{\theta}_{\mathrm{S}}$ and $\cos\bar{\theta}_{\mathrm{L}}$ uniformly distributed in the range $[-1,1]$ and $\bar{\phi}_{\mathrm{S}}$ and $\bar{\phi}_{\mathrm{L}}$ in the range $[0,2\pi]$.
Then we calculate SNR of each binary, fixing the redshifted NS masses at $1.4 (1+z) M_{\odot}$.
We denote the SNR of $i$-th binary out of the $10^4$ binaries at redshift $z$ as $\rho(z,i)$.
We start integrations of GW signals  from $f_{\mathrm{min}}=0.2$Hz, considering the potential confusion noise by WD/WD binaries.
We set the upper  cutoff frequency  $f_{\mathrm{max}}=100$Hz, but our results are insensitive to this choice.

For the detector configurations, we consider 4 cases shown in Fig.~\ref{config-cases} with DECIGO.
To concentrate on the geometrical effects, we use the fixed number  of the triangle units (totally four, corresponding to  eight L-shaped interferometers). These units  are assumed to have the identical sensitivity to DECIGO, and only their positions and orientations  are different. Note that our analysis  is based on the long-wave approximation for responses of detectors, and  the two units forming a star-of-David constellation can be regarded as a completely aligned pair. 
The results below remain almost the same when we use BBO instead (as we confirm in Sec.~\ref{sec-res}).

\subsection{Minimum SNR}

\begin{figure}[tbp]
  \includegraphics[scale=.7,clip]{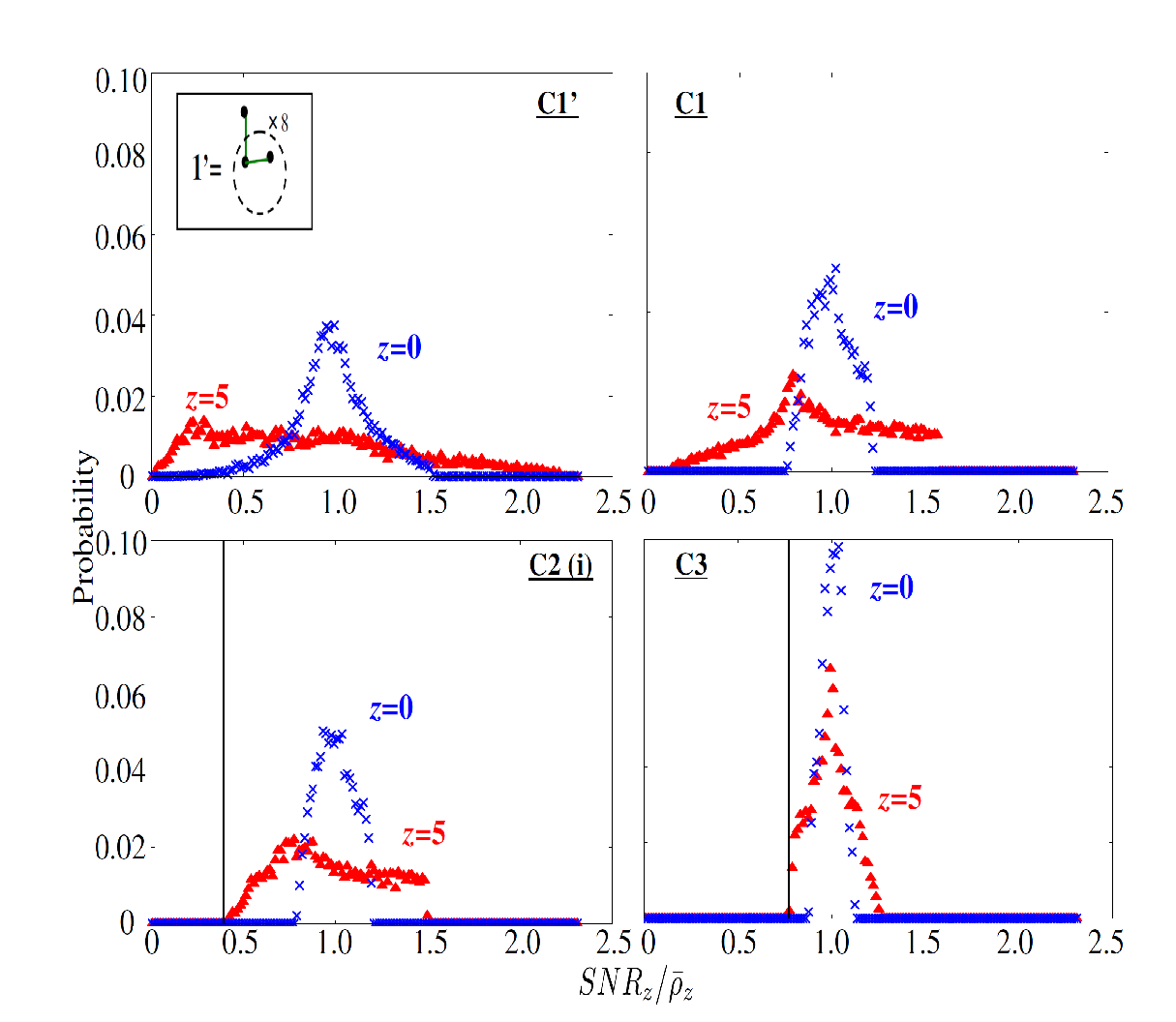} 
 \caption{\label{z0z5} These figures show the probability distributions of binaries in each SNR bins, comparing the results for both binaries at $z=0$ and $z=5$.
We normalize each SNR with the root-mean-square value $\bar{\rho}_z$.
Each panel on top left, top right, bottom left and bottom right corresponds to the one using detectors of the configurations C1', C1, C2(i) and C3, respectively.
Here, C1' represents the configuration with 8 right-angled interferometers placed on one site, as shown in the small panel on top left.
The (red) triangular plots in each panel show the results for binaries at $z=5$ while the (blue) crosses indicate the ones at $z=0$.
The solid line in each panel of C2(i) and C3 represents the value $G$ for each optimal configuration with a static detector network.
The dashed line in C3 panel shows the one $G=0.895$ with infinite number of detectors (see Sec.~\ref{three}). }
\end{figure}

In Fig.~\ref{z0z5}, we  show  the probability distributions of SNRs for the edge-on NS/NS binaries at $z=0$ and 5.  
We count the number of binaries within each SNR bin and divide it by the total number of binaries 10$^4$.
In order to compare the overall distribution pattern of SNRs  at the two redshifts, we normalize the SNRs by $\bar{\rho}_z$.
We consider the configurations C1, C2(i) and C3, with the latter two at the optimal separation angles ($\alpha_2=52.4^{\circ}$, $\alpha_3=126.36^{\circ}$).
For reference, we also consider a somewhat extreme configuration denoted as C1'.
This has totally eight L-shaped interferometers placed on one site. 

First, let us look at the results of $z=5$.
We see that as we increase the number of detectors, the distributions become sharper due to the averaging effect of the beam pattern functions.
For the distributions of C1 and C1', there are tails on lower SNR side, down to $SNR_5/\bar{\rho}_5 \approx 0$.
This is because there exist dead angles of binaries for the static detectors having only one detector plane.
However, since the detectors are moving during the observation period of about 3 weeks, the minimum SNR for each configuration is not exactly 0. 
On the other hand, there exists a distinct cutoff on lower SNR side (and also on higher SNR side) for the configurations  C2(i) and C3.  
This is because these networks with more than 1 detector plane do not have dead angle directions even for the static case.
This cutoff represents the minimum SNR (normalized with $\bar{\rho}_5$) for all NS/NS binaries detected by DECIGO/BBO and therefore corresponds to the value $G$ in Sec.~\ref{sec-dead} {but now  for the case of moving detector networks}.
If the detector sensitivity is high enough to identify the signal of this lowest SNR binary,  we can handle  individual contributions of all binaries in the data streams of detectors (provided that, ultimately, subtraction of individual binaries are not limited by the  NS/NS foreground itself and the residual noises after the binary subtraction would be sufficiently small). 
In each panel of C2(i) and C3 in Fig.~\ref{z0z5}, we show the analytical value of $G$ {for the static detector network} as a solid line, which is obtained in Sec.~\ref{sec-dead}.
We can see that this analytical value almost matches with the lower cutoff value obtained by the Monte Carlo simulation. 
Again, due to the motions of detectors, the numerical counterparts are slightly larger than the ones estimated analytically for the static configurations.

Next, we compare the results of $z=5$ and $z=0$.
Figure~\ref{z0z5} shows that the SNR probability distributions of $z=0$ are sharper than the ones of $z=5$.
This is because the observation time for $z=0$ is longer than for $z=5$, and the number of independent detector planes are effectively increased. 
In the panel of configuration C3, we also show the analytic value $G=\sqrt{205}/16=0.895$ given for infinite number of detectors as a dashed line.
This value almost matches with the lower cutoff value of $z=0$ in the same panel.
Since observation time of binaries at $z=0$ is slightly larger than 1 yr, detectors orbit around the Sun completely.
Therefore, as we increase the number of detector planes, the lower cutoff in the probability distribution approaches to $G=0.895$.

 \begin{figure}[tbp]
  \includegraphics[scale=.7,clip]{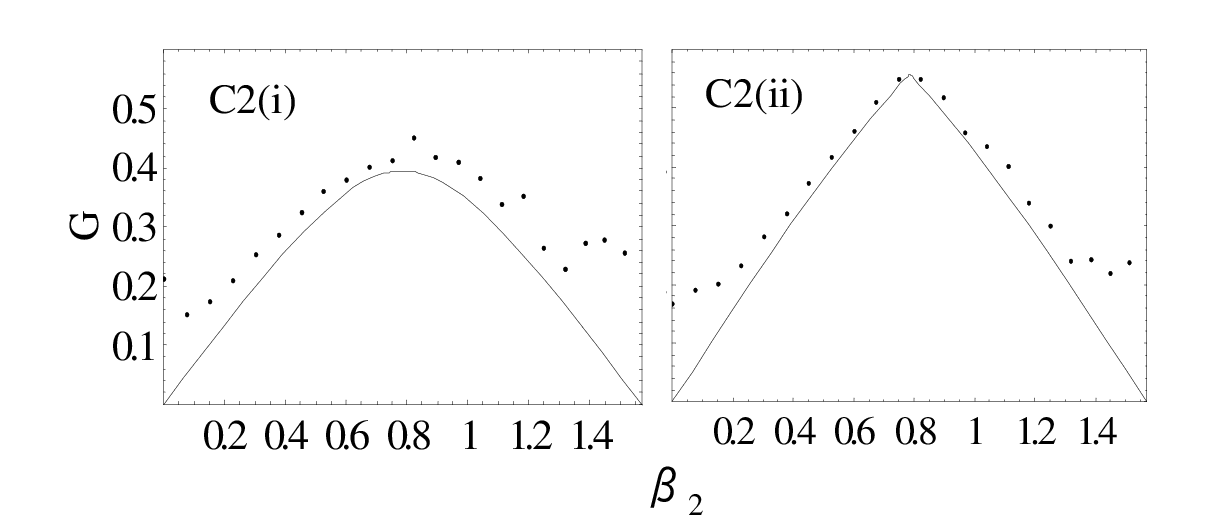} 
 \caption{\label{opt2d-num} 
The normalized ratio $G_5\equiv {SNR_{5,\mathrm{min}}}/{\bar{\rho}_5} $ for C2(i) and C2(ii) as functions of the opening angle $\beta_2$.
The solid line represents the analytical results shown in Fig.~\ref{opt2d} assuming that detectors are static.
The dots are the ones obtained by our numerical calculations which include the effects of detector motions. }
\end{figure}

\begin{figure}[tbp]
  \includegraphics[scale=.5,clip]{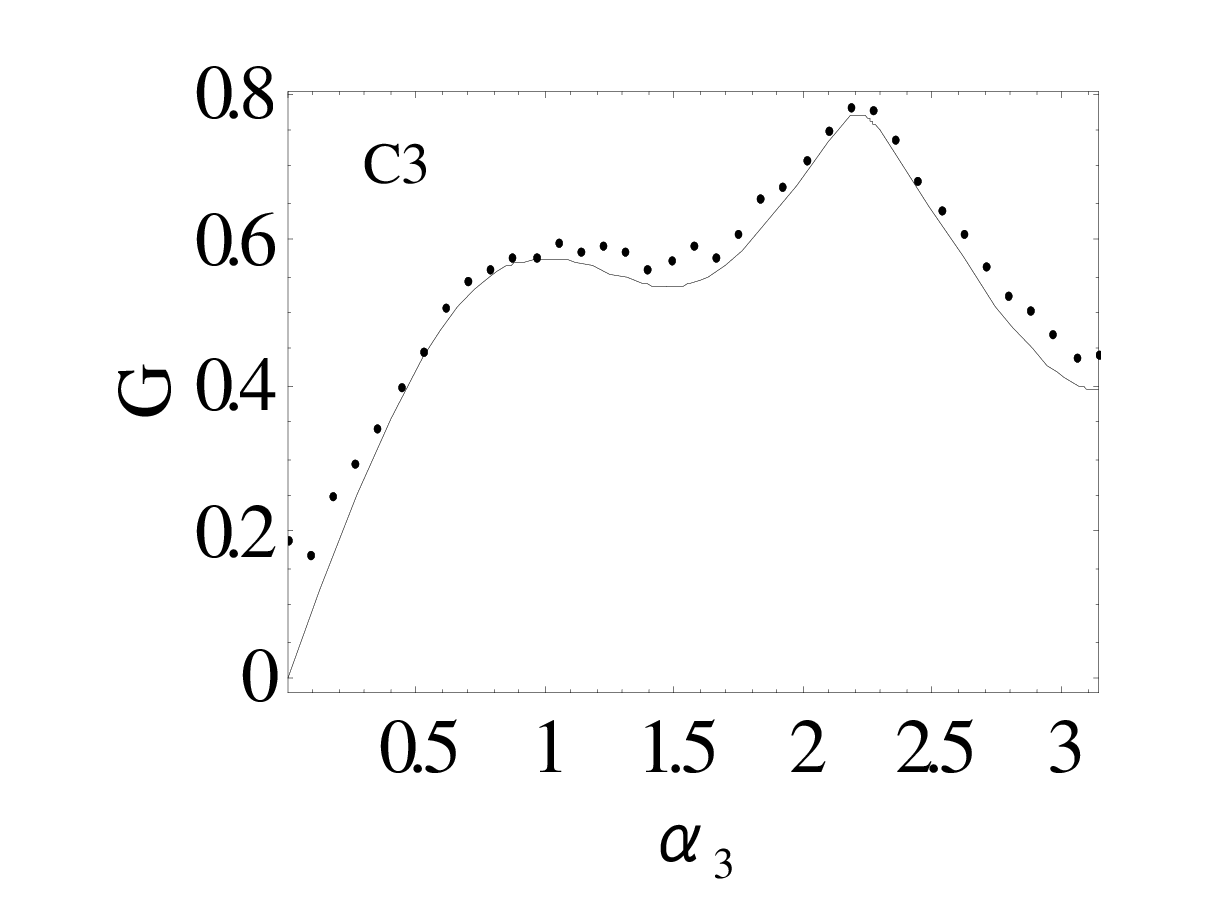} 
 \caption{\label{opt3d-num}
The normalized ratio $G_5\equiv {SNR_{5,\mathrm{min}}}/{\bar{\rho}_5} $ for C3 as a function of the detector separation angle $\alpha_3$. 
The meanings for the line and dots are the same as in Fig.~\ref{opt2d-num}.}
\end{figure}

Now, we compare the analytical results of the values $G$ for the static detectors shown in Figs.~\ref{opt2d} and \ref{opt3d} with our numerical results which include the effects of detector motions.
In Fig.~\ref{opt2d-num}, we present $G_5$ for the configurations C2(i) and C2(ii) that have only two detector planes.
Solid line on each panel indicates the analytical result, while the dots are the ones obtained by our numerical simulations.
We can see that the analytical and numerical results are quite close for the cases of large $G_5$, while they do not match for the cases of small $G_5$, especially where $G_5\approx 0$.
This shows that the effects of detector motions are more significant for the smaller $G_5$ cases.
When detectors are static, there exist dead angles for the configurations $\beta_2=0$ and $\beta_2=\pi /2$.
These dead angles disappear when detectors are moving, which makes $G_5 \neq$ 0 and causes a remarkable difference between the static and the moving cases.
Also, numerical results are not symmetric with respect to $\beta_2 =\pi /4$, as discussed in Sec. V.B.
For the case of $\beta_2=0$, the number of detector plane becomes one and the results for C2(i) and C2(ii) should match.
The discrepancy comes from the statistical fluctuations which are caused by the finiteness of the  number of binaries in our Monte Carlo simulations.

In Fig.~\ref{opt3d-num}, we show the results for the case C3.
In this case, the distinction between analytical and numerical results is smaller than the cases C2.
The maximum value of $G_5$ is obtained when $\alpha_3$ is around 126$^\circ$, though it is almost the same with the one for the default configuration of $\alpha_3$=120$^{\circ}$.
From Figs.~\ref{opt2d-num} and~\ref{opt3d-num}, we understand that the matching between the analytical and numerical results becomes better as we increase the symmetry of detector configurations.
This tendency is consistent with the discussion made in Sec.~\ref{sec-dead}.
As we increase the symmetry, the effect of detector motions is suppressed and in the limit of infinite number of detector configuration, the values of $G_5$ are exactly the same for the static and moving cases.

Next we evaluate the network sensitivity $\bar{\rho}_5$ required for identifying all the binaries with our fiducial merger model in which NS/NS signals exist up to $z=5$.  With respect to the minimum value $SNR_{5,\mathrm{min}}$ of our binaries, we have
\beq
G_5=\frac{SNR_{5,\mathrm{min}}}{\bar{\rho}_5}
\eeq
due to the definition of the parameter $G_5$.
For complete identification of the binaries, we should have 
\beq
{SNR_{5,\mathrm{min}}}>\rho_{\mathrm{thr}}
\eeq
with the detection threshold $\rho_{\mathrm{thr}}$ (which may depend on the future computing power and the number of templates required to cover the parameter space~\cite{cutlerharms}).
Then we  get
\beq
\frac{\rho_{\mathrm{thr}}}{\bar{\rho}_5}<G_5
\eeq
or equivalently
\beq
{\bar{\rho}_5}>G_5^{-1}{\rho_{\mathrm{thr}}}=25.9  \lmk\frac{G_5}{0.77}  \rmk^{-1} \lmk  \frac{\rho_{\mathrm{thr}}}{20} \rmk, \label{ineq}
\eeq
where we used the parameter $G_5=0.77$ for the case C3 and the typical threshold $\rho_{\mathrm{thr}}=20$.
With the network sensitivity of $\bar{\rho}_5=10.4$ for DECIGO, many binaries have SNRs below   $\rho_{\mathrm{thr}}=20$ and would not be identified.  To fix this, the sensitivity of DECIGO should be improved  by a factor of $2.5 \lmk \frac{\rho_{\mathrm{thr}}}{20} \rmk$.
In contrast, BBO has  $\bar{\rho}_5=33.6$ and,  with $G_5=0.77$, it can detect all the NS/NS signals with SNRs higher  than 33.6$\times$0.77=25.9.
However, for a higher threshold $\rho_{\mathrm{thr}}=30$, the network  sensitivity is not sufficient even for BBO and then we   are not able to perform complete subtraction.
In the next subsection, we study  what fraction of them remains to be unidentified.
Notice that the results in this subsection are not affected by the values of $\dot{n}_0$.

\begin{figure}[tbp]
  \includegraphics[scale=.8,clip]{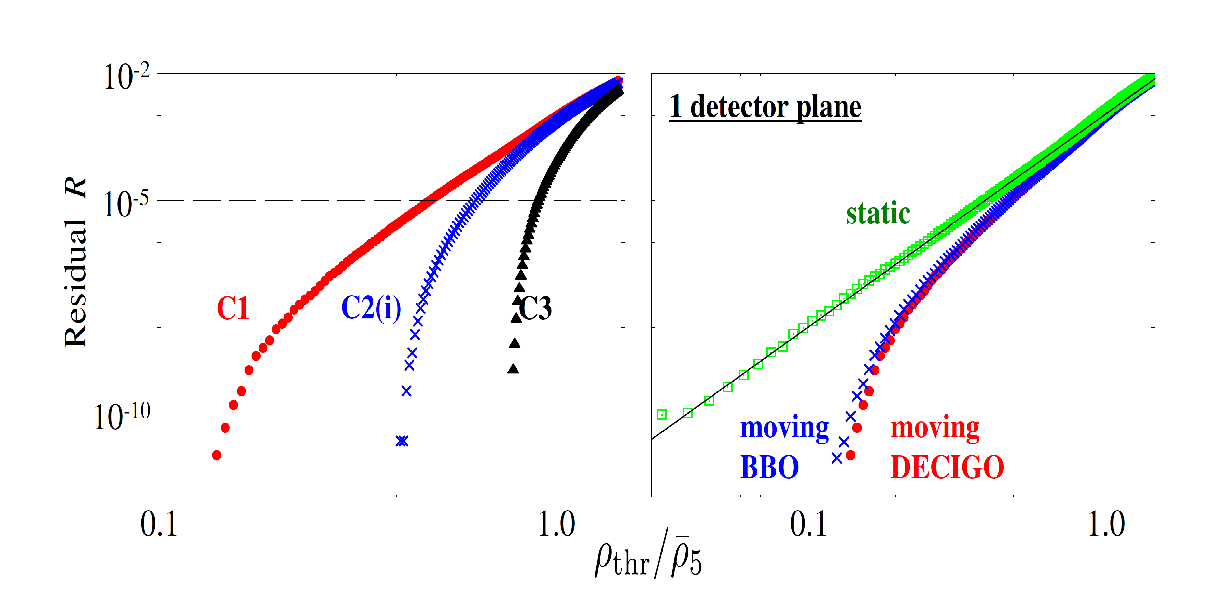} 
 \caption{\label{snthr} 
(Left) This figure shows the accumulated fractional residuals $R$ against $\rho_{\mathrm{thr}}/\bar{\rho}_5$ for various configurations.
The (red) circular plots, the (blue) crosses and the (black) triangular plots show the ones using detectors of configuration C1, C2(i) and C3, respectively.
For each case of C2(i) and C3, we take the optimal configuration.
We assume that the subtraction is succeeded if the residual is below the horizontal dashed line of $R=10^{-5}$.
(Right) This figure also shows the residuals $R$ for the detectors having only 1 detector plane.
The (red) circular plots are the results using the moving DECIGO and this is same as the one shown in the left panel.
The (blue) crosses indicate the ones using the moving BBO while the (green) square plots are the ones using the static DECIGO, with the fitted line of $R\propto (\rho_{\mathrm{thr}}/\bar{\rho}_5)^5$ shown as the black solid line.}
\end{figure}

\subsection{Residuals}
\label{sec-res}

In order to estimate the amplitude  of unidentified foreground composed by binaries with SNRs less than the threshold $\rho_{\mathrm{thr}}$, we  define the ratio $\gamma (z)$ for the numerical samples described in Sec. VI.B as
\begin{equation}
\gamma (z) \equiv \frac{\sum_i \rho(z,i)^2 \Theta [\rho_{\mathrm{thr}}-\rho(z,i)]}{\sum_i \rho(z,i)^2} \label{gamma}
\end{equation}
with the step function $\Theta(\cdot)$.
Then, from Eq.~(\ref{omega-NS}), we can calculate the fraction of the residual NS/NS binaries in the total foreground $\Omega_{\mathrm{GW}}^{\mathrm{NS}}$ as
\begin{equation}
R\equiv \frac{\int _{0}^{\infty} dz \gamma (z) \frac{s(z)}{(1+z)^{4/3}H(z)}}{\int _{0}^{\infty} dz \frac{s(z)}{(1+z)^{4/3}H(z)}}.
\end{equation}
The basic question here is how the residual fraction $R$ depends on the configuration of detectors.

In the left panel of Fig.~\ref{snthr}, we show the residual $R$ against $\rho_{\mathrm{thr}}/\bar{\rho}_5$ for the cases C1, C2(i) and C3.
For the case of C3, the residual falls down to 0 very steeply as we lower $\rho_{\mathrm{thr}}/\bar{\rho}_5$.
This figure shows that in order to achieve $R<10^{-5}$ for our fiducial merger rate of $\dot{n}_0=10^{-7}$ Mpc$^{-3}$ yr$^{-1}$, the network sensitivity $\bar{\rho}_5$ should satisfy $\rho_{\mathrm{thr}}/\bar{\rho}_5<0.91$.
However, since the value of $R$ depends very sensitively on $\rho_{\mathrm{thr}}/\bar{\rho}_5$, we suggest that it seems much safer to aim for $\rho_{\mathrm{thr}}/\bar{\rho}_5< 0.77$ so that all of the binary signals can be identified.
For the case of C2(i), the residual $R$ depends more gently on $\rho_{\mathrm{thr}}/\bar{\rho}_5$ and $R<10^{-5}$ requires $\rho_{\mathrm{thr}}/\bar{\rho}_5<0.63$.
In this case, it might be reasonable to adjust the network sensitivity $\bar{\rho}_5$ to perform successful subtraction down to $R<10^{-5}$.
However with the sensitivity $\rho_{\mathrm{thr}}/\bar{\rho}_5<0.63$, we suggest to use the C3 configuration and  subtract all the binary signals.
For the case of C1, we need a better sensitivity $\rho_{\mathrm{thr}}/\bar{\rho}_5<0.48$ to accomplish $R<10^{-5}$, which seems a rather demanding  task.

Next, we compare the results for DECIGO and BBO to check that they give almost the same results.
Out of the configurations C1, C2 and C3, it is expected that the configuration C1 gives the largest discrepancy between these two results.
On the right panel of Fig.~\ref{snthr}, we show the dependence of residuals on $\rho_{\mathrm{thr}}/\bar{\rho}_5$ using detectors having only one detector plane (C1).
The (red) circular plots are the ones using DECIGO and this is the same as the ones on the left panel, while the (blue) crosses represents the ones using BBO.
Since the results for the moving DECIGO and BBO are almost the same, we can confirm that our main results obtained in this paper can be applied to BBO as well.

On the same panel, we show the residuals obtained by using the static DECIGO and also provide a fitted line of $R\propto  (\rho_{\mathrm{thr}}/\bar{\rho}_5)^{5}$.
This index of $5$ can be understood analytically, as follows.
When considering detectors with one detector plane, we have the  dead angles $(\theta_{\mathrm{S}},\psi_{\mathrm{S}},i)=(\pi /2,\pi /4,\pi /2)$ and we expect that the  unidentified binaries have  angular parameters similar to these specific combinations.
Thus,  around the dead angles, we introduce the coordinate system ($\Delta\theta_{\mathrm{S}}, \Delta\psi_{\mathrm{S}}, \Delta i$), and  consider the region   where the SNR is less than a given threshold  $\rho_{\mathrm{thr}}$.
From Eq.~(\ref{dead}),  the region turns out to be  a sphere, and its radius is proportional to $\rho_{\mathrm{thr}}$.
Therefore,  the number of binaries $N_{\mathrm{<thr}}$ within the region has the following relation     
\begin{equation}
N_{\mathrm{<thr}} \propto \Delta\theta_{\mathrm{S}} \Delta\psi_{\mathrm{S}} \Delta i \propto \rho_{\mathrm{thr}}^3. \label{nthr}
\end{equation}
Since the averaged SNR in the region is proportional to $\rho_{\mathrm{thr}}$, together with Eq.~(\ref{nthr}), the unidentified fraction defined in Eq.~(\ref{gamma}) scales as
\begin{equation}
\gamma \propto \rho_{\mathrm{thr}}^5.
\end{equation} 
Since the fractional residuals $R$ is roughly proportional to $\gamma$, we can reproduce the simple dependence $R \propto (\rho_{\mathrm{thr}}/\bar{\rho}_5)^5$. 
When we include the detector motions, the residual starts to deviate from this power-law as we lower the value of $\rho_{\mathrm{thr}}/\bar{\rho}_5$.

\section{CONCLUSIONS}
\label{sec-conclusion}

In this paper, we estimated the detector  noise level  required  to  individually identify foreground GWs made by cosmological NS/NS binaries. We characterized the total sensitivity of the detector network by an integral ${\bar{\rho}_5}$, assuming that the binaries exist at $z\le 5$.
We mainly focused on the relationship between the geometrical properties of the detectors and the  minimum SNR of the binaries, and introduced a convenient parameter $G_z$ defined by
\beq
G_z\equiv\frac{SNR_{z,\mathrm{min}}}{\bar{\rho}_z}\le 1. 
\eeq

We started our studies  with static detector networks. 
In contrast to a network with a single detector plane, in general the minimum SNR of binaries becomes finite for the one with more than 1 detector planes.
For the C3 configuration (see Fig.~\ref{default}) with 3 detector planes, we analytically estimated the optimal angle $\alpha_3=126.4^\circ$  with $G_z=0.771$.
It is the angles between detector planes that play important roles in realizing a high $G$ and not the separations between detector units (although they are important for improving the angular resolution).

We expected that the ratio $G_z$ becomes larger when we use the moving detectors, due to an averaging effect.
In order to  estimate this correction, we performed  Monte Carlo simulations, randomly distributing cosmological NS/NS binaries at $z\le 5$. 
We showed that our analytical predictions and numerical results match well, especially for large $G_5$ cases.
In order to identify all of NS/NS binaries, we need to achieve enough detector sensitivity so that the  minimum SNR  is larger than the detection threshold $\rho_{\mathrm{thr}}$.
This leads to the following inequality 
\beq
{\bar{\rho}_5}>G_5^{-1}{\rho_{\mathrm{thr}}}=25.9  \lmk\frac{G_5}{0.77}  \rmk^{-1} \lmk  \frac{\rho_{\mathrm{thr}}}{20} \rmk.
\eeq
It suggests that for  a given threshold at  $\rho_{\mathrm{thr}}=20$,  all the binary signals can be identified if the  optimal C3 network has a sensitivity  $\bar{\rho}_5>25.9$.  With its reference parameters, DECIGO has the network sensitivity  $\bar{\rho}_{5}=10.4$.  In order to perform complete identification of NS/NS binaries at $z\le 5$, its sensitivity should be improved by a factor of $2.5 \lmk \frac{\rho_{\mathrm{thr}}}{20} \rmk$.  BBO  has $\bar{\rho}_{5}=33.6$ and can identify all of them as long as $\rho_{\mathrm{thr}}=20$.
However, for a higher threshold  $\rho_{\mathrm{thr}}=30$, even BBO fails to subtract all the binary signals.
At the end of our calculation, we estimated the residual fraction $R$ of the foreground that cannot be identified.
We found that the low SNR end of the residual is very steep.  Therefore we  recommended  to realize  the sensitivity sufficient for complete identification rather than those with $R<10^{-5}$.

When calculating SNR in this paper, we only used the instrumental noise spectral density.
In order to perform more realistic analysis, we need to include NS/NS foreground noises and perform self-consistent analysis as done in Ref.~\cite{cutlerharms}.
Furthermore, although   the individual GW signals are assumed to be removed once they have been identified, this is not a straightforward step.  With respect to this issue, we should perform more elaborate data analysis simulations (see {\it e.g.} Ref.~\cite{harms} and also Refs.~\cite{babak,crowder}).

In this paper, we have assumed circular binaries with no spins.
Cutler and Harms~\cite{cutlerharms} discussed that typical NS/NS binaries have squared eccentricities $e^2\sim 10^{-8}$ at $f=0.3$Hz.
For these binaries, the correction in the GW phase due to eccentricities become $\sim$ 0.2 for the dominant harmonic, and we should  include this phase effect in the template for matched filtering.
They also showed that the spin-orbit correction to the phase becomes important when one of the binary component has spin period $\sim$10 msec.
However, even in this case, the contributions from spin-spin couplings and precessions can be neglected. 

It is expected that we can strongly constrain cosmological parameters by observing GWs from chirping binaries.  
The crucial point here  is whether we can  determine the redshifts $z$ of the GW sources  by associated electro-magnetic wave observations (see Cutler and Holz~\cite{cutlerholz} and references therein).
It might be useful to study dependence of  the angular resolution of binaries on configurations of detectors.  Another aspect closely related to the geometry of the detector  network is the decomposition of various polarization modes of GWs, including parity asymmetry (the Stokes $V$ parameter) \cite{Seto:2006dz} and the non-Einstein modes \cite{nishizawa}.

\begin{acknowledgments}
\quad We thank Takashi Nakamura and Takahiro Tanaka for useful discussions, and the referees for valuable comments.
We also thank Seiji Kawamura, Kenji Numata and Tomotada Akutsu for giving us information on DECIGO sensitivities.
K.Y. is supported by the Japan Society for the Promotion of Science grant No. $22 \cdot 900$ and N.S. is supported by the Ministry of Education, Culture, Sports, Science and Technology (MEXT) of Japan grant No. 20740151.
This work is also supported in part by the Grant-in-Aid for the Global COE Program ``The Next Generation of Physics, Spun from Universality and Emergence'' from the MEXT of Japan. 
\end{acknowledgments}

\appendix

\section{Useful formulae }
\label{app}

When we perform Monte Carlo simulations, we randomly distribute the direction of the source $(\bar{\theta}_{\mathrm{S}},\bar{\phi}_{\mathrm{S}})$ and the direction of the orbital angular momentum $(\bar{\theta}_{\mathrm{L}},\bar{\phi}_{\mathrm{L}})$, both measured in the solar barycentric frame.
Therefore we need to express the waveforms  in terms of $\bar{\theta}_{\mathrm{S}},\bar{\phi}_{\mathrm{S}},\bar{\theta}_{\mathrm{L}}$ and $\bar{\phi}_{\mathrm{L}}$ (see \cite{cutler1998} for details). The angles 
$\theta_{\mathrm{S}}(t)$ and $\phi_{\mathrm{S}}(t)$ in the moving detector frame are expressed as
%
\begin{eqnarray}
\cos \theta_{\mathrm{S}}(t)&=&\frac{1}{2}\cos \bar{\theta}_{\mathrm{S}}
                                          -\frac{\sqrt{3}}{2}\sin \bar{\theta}_{\mathrm{S}} \cos[\bar{\phi}(t)-\bar{\phi}_{\mathrm{S}}],  \\
\phi_{\mathrm{S}}(t)&=&\frac{\pi}{12}+\tan^{-1}
                              \left( \frac{\sqrt{3}\cos{\bar{\theta}_{\mathrm{S}}}
                              +\sin \bar{\theta}_{\mathrm{S}}\cos [\bar{\phi}(t)-\bar{\phi}_{\mathrm{S}}]}
                              {2\sin \bar{\theta}_{\mathrm{S}}\sin [\bar{\phi}(t)-\bar{\phi}_{\mathrm{S}}]} \right).
\end{eqnarray}
%
The polarization angle $\psi_{\mathrm{S}}$ in the moving detector frame is given by Eq.~(\ref{tanpsi}) with the following three quantities
$\hat{\bm{L}}\cdot\hat{\bm{z}}$, $\hat{\bm{L}}\cdot\hat{\bm{N}}$ and $\hat{\bm{N}}\cdot(\hat{\bm{L}}\times\hat{\bm{z}})$. Here we have 
\begin{eqnarray}
\hat{\bm{L}}\cdot\hat{\bm{z}}&=&\frac{1}{2}\cos\bar{\theta}_{\mathrm{L}}
                   -\frac{\sqrt{3}}{2}\sin\bar{\theta}_{\mathrm{L}} \cos[\bar{\phi}(t)-\bar{\phi}_{\mathrm{L}}], \label{lz} \\
\hat{\bm{L}}\cdot\hat{\bm{N}}&=&\cos\bar{\theta}_{\mathrm{L}}\cos\bar{\theta}_{\mathrm{S}}
                  +\sin\bar{\theta}_{\mathrm{L}}\sin\bar{\theta}_{\mathrm{S}}\cos(\bar{\phi}_{\mathrm{L}}-\bar{\phi}_{\mathrm{S}}), \label{ln} \\
\hat{\bm{N}}\cdot(\hat{\bm{L}}\times\hat{\bm{z}})&=&\frac{1}{2}\sin\bar{\theta}_{\mathrm{L}}\sin\bar{\theta}_{\mathrm{S}}
                                             \sin(\bar{\phi}_{\mathrm{L}}-\bar{\phi}_{\mathrm{S}}) \notag \\
                                             &&-\frac{\sqrt{3}}{2}\cos\bar{\phi}(t)(
                                             \cos\bar{\theta}_{\mathrm{L}}\sin\bar{\theta}_{\mathrm{S}}\sin \bar{\phi}_{\mathrm{S}} 
                                             -\cos\bar{\theta}_{\mathrm{S}}\sin\bar{\theta}_{\mathrm{L}}\sin \bar{\phi}_{\mathrm{L}} ) \notag \\
                                             &&-\frac{\sqrt{3}}{2}\sin\bar{\phi}(t)(
                                             \cos\bar{\theta}_{\mathrm{S}}\sin\bar{\theta}_{\mathrm{L}}\cos \bar{\phi}_{\mathrm{L}} 
                                             -\cos\bar{\theta}_{\mathrm{L}}\sin\bar{\theta}_{\mathrm{S}}\cos \bar{\phi}_{\mathrm{S}} ). \label{nlz}
\end{eqnarray}




\end{document}